\documentclass{aa}

\usepackage{graphicx}
\usepackage{float}
\usepackage{amsmath}
\usepackage{amssymb}
\usepackage{natbib}
\usepackage{xcolor}
\usepackage{import}
\usepackage[varg]{txfonts}

\usepackage{enumitem}

\def\aap{A\& A}
\def\apj{ApJ}
\def\mnras{MNRAS}
\def\aj{AJ}
\def\aaps{AAPS}

\def\araa{ARA\& A}
\def\aapr{A\& ARv}

\usepackage[T1]{fontenc}
\usepackage{ae,aecompl}
\usepackage{hyperref}

\newcommand{\degree}{^{\circ}}

\begin{document}

\title{The Strong Gravitational  Lens Finding Challenge}

\author{
R. Benton Metcalf \inst{1,2}\thanks{E-mail: robertbenton.metcalf@unibo.it} 
\and
M. Meneghetti \inst{2}
\and
Camille Avestruz \inst{3,4,5}\fnmsep\thanks{Provost's Postdoctoral Scholar at the University of Chicago}
\and
Fabio Bellagamba\inst{1,2}
\and
Cl\'ecio R. Bom\inst{6,7}
\and
Emmanuel Bertin\inst{8}
\and
R\'emi Cabanac \inst{9}
\and
F. Courbin \inst{12}
\and
Andrew Davies \inst{22}
\and
Etienne Decenci\`ere \inst{10}
\and
R\'emi Flamary \inst{11}
\and
Raphael Gavazzi \inst{8}
\and
Mario Geiger \inst{12} 
\and
Philippa Hartley \inst{13}
\and
Marc Huertas-Company \inst{14}
\and
Neal Jackson \inst{13}
\and
Colin Jacobs\inst{23}
\and
Eric Jullo \inst{15}
\and
Jean-Paul Kneib \inst{12}
\and
L\'{e}on V. E. Koopmans \inst{16}
\and
Fran\c{c}ois Lanusse \inst{17}
\and
Chun-Liang Li \inst{18}
\and
Quanbin Ma \inst{18}
\and
Martin Makler \inst{7}
\and
Nan Li \inst{19}
\and
Matthew Lightman \inst{15}
\and
Carlo Enrico Petrillo \inst{16}
\and
Stephen Serjeant \inst{22}
\and
Christoph Sch\"{a}fer \inst{12}
\and
Alessandro Sonnenfeld \inst{21}
\and
Amit Tagore \inst{13}
\and
Crescenzo Tortora \inst{16}
\and
Diego Tuccillo \inst{10,14}
\and
Manuel B. Valent\'in \inst{7}
\and
Santiago Velasco-Forero \inst{10}
\and
Gijs A. Verdoes Kleijn \inst{16}
\and
Georgios Vernardos \inst{16}
}

\institute{
Dipartimento di Fisica \& Astronomia, Universit\`a di Bologna, via Gobetti 93/2, 40129 Bologna, Italy 
\and
INAF-Osservatorio Astronomico di Bologna, via Ranzani 1, 40127 Bologna, Italy 
\and
Enrico Fermi Institute, The University of Chicago, Chicago, IL 60637 U.S.A.
\and
Kavli Institute for Cosmological Physics, The University of Chicago, Chicago, IL 60637 U.S.A.
\and
Department of Astronomy \& Astrophysics, The University of Chicago, Chicago, IL 60637 U.S.A.
\and
Centro Federal de Educa\c{c}\~ao Tecnol\'ogica Celso Suckow da Fonseca, CEP 23810-000,  Itagua\'i, RJ, Brazil
\and
Centro Brasileiro de Pesquisas F\'isicas, CEP 22290-180, Rio de Janeiro, RJ, Brazil
\and
Institut d'Astrophysique de Paris, Sorbonne Universit\'e, CNRS,  UMR 7095, 98 bis bd Arago, 75014 Paris, France.
\and
IRAP, Universit\'e de Toulouse, CNRS, UPS, Toulouse, France.
\and
MINES Paristech, PSL Research University, Centre for Mathematical Morphology, 35 rue  Saint-Honore, Fontainebleau, France
\and
Laboratoire Lagrange, Universi\'{e} de Nice Sophia-Antipolis, Centre National de la Recherche Scientifique,
\and
Institute of Physics, Laboratory of Astrophysics, Ecole Polytechnique F\'ed\'erale de Lausanne (EPFL),  Observatoire de Sauverny, 1290 Versoix, Switzerland 
\and
Jodrell Bank Centre for Astrophysics, School of Physics \& Astronomy, University of  Manchester, Oxford Rd, Manchester M13 9PL, UK Observatoire de la C\^{o}te d'Azur, Parc Valrose, 06108 Nice, France 
\and
LERMA, Observatoire de Paris, CNRS, Universit\'e Paris Diderot, 61, Avenue de l'Observatoire F-75014, Paris, France
\and
Aix Marseille Universit\'e, CNRS, LAM (Laboratoire d'Astrophysique de Marseille) UMR 7326, 13388, Marseille, France
\and
Kapteyn Astronomical Institute, University of Groningen, Postbus 800, 9700 AV, Groningen, The Netherlands
\and
McWilliams Center for Cosmology, Department of Physics, Carnegie Mellon University, Pittsburgh, PA 15213, USA
\and
School of Computer Science, Carnegie Mellon University, Pittsburgh, PA 15213, USA
\and
School of Physics and Astronomy, Nottingham University, University Park, Nottingham, NG7 2RD, UK
\and
JPMorgan Chase, Chicago, IL 60603 U.S.A.
\and
Kavli IPMU (WPI), UTIAS, The University of Tokyo, Kashiwa, Chiba 277-8583, Japan
\and
School of Physical Sciences, The Open University, Walton Hall, Milton Keynes, MK7 6AA, UK
\and
Centre for Astrophysics and Supercomputing, Swinburne University of Technology, P.O. Box 218, Hawthorn, VIC 3122, Australia
}

\abstract{
Large-scale imaging surveys will increase the number of galaxy-scale strong lensing candidates by maybe three orders of magnitudes beyond the number known today.  Finding these rare objects will require picking them out of at least tens of millions of images, and deriving scientific results from them will require quantifying the efficiency and bias of any search method.  To achieve these objectives automated methods must be developed.
Because gravitational lenses are rare objects, reducing false positives will be particularly important.  We present a description and results of an open gravitational lens finding challenge.  Participants were asked to classify 100,000 candidate objects as to whether they were gravitational lenses or not with the goal of developing better automated methods for finding lenses in large data sets.   A variety of methods were used including visual inspection, arc and ring finders, support vector machines (SVM) and convolutional neural networks (CNN).  We find that many of the methods will be easily fast enough to analyse the anticipated data flow.  In test data, several methods are able to identify upwards of half the lenses after applying some thresholds on the lens characteristics such as lensed image brightness, size or contrast with the lens galaxy without making a single false-positive identification.  This is significantly better than direct inspection by humans was able to do.  Having multi-band, ground based data is found to be better for this purpose than single-band space based data with lower noise and higher resolution, suggesting that multi colour data is crucial.  Multi-band space based data will be superior to ground based data.
The most difficult challenge for a lens finder is differentiating between rare, irregular and ring-like face-on galaxies and true gravitational lenses.  The degree to which the efficiency and biases of lens finders can be quantified largely depends on the realism of the simulated data on which the finders are trained.
}

\maketitle

\section{Introduction}
\label{sec:introduction}

Strong gravitational lenses are rare cases in which a distant galaxy or quasar is aligned so closely with a foreground galaxy or cluster of galaxies  that the gravitational field of the foreground object creates multiple, highly distorted images of the background object.   The first strong lens was discovered in 1979 by \citet{1979Natur.279..381W} and since then several hundred of them have been found.
When the lens is an individual galaxy and the source a quasar, there are two, four or five distinct images of the source.  The time-delay between images,  the magnification ratios between images and the image positions can all be used to model the mass distribution of the lens and measure cosmological parameters.  When the lens is a cluster of galaxies the images of background galaxies are multiplied and distorted into many thin arcs.   When the lens is an individual galaxy and the background source also a galaxy, the lensed images can take the form of a partial or complete ring seen around or through the lens galaxy, an {\it Einstein ring}.   

Strong lenses have to date provided very valuable scientific information.  They have been used to study how dark matter is distributed in galaxies and clusters \citep[e.g.][]{1991ApJ...373..354K,2001ApJ...554.1216C,2002ApJ...568L...5K,2003ApJ...587..143R,2003ApJ...583..606K,2005MNRAS.360.1333W,2005ApJ...623...31D,2009MNRAS.392..945V,2016MNRAS.463.3115T} and  to measure the Hubble constant and other cosmological parameters 
\citep[e.g.][]{1964MNRAS.128..307R,1992ARAandA..30..311B,2000ApJ...544...98W,2013ApJ...766...70S,2016AandARv..24...11T}.  
Their magnification has been used as a natural telescope to observe otherwise undetectable objects at high redshift \citep[e.g.][]{2007ApJ...671.1196M,2017MNRAS.464.4823B,2016ApJ...833..264S}.   They have put limits on the self interaction of dark matter and on the theory of gravitation \citep{0004-637X-606-2-819}.  Through microlensing they have been used to study the structure of quasars \citep{2008ApJ...689..755M,2008ApJ...673...34P,2011ApJ...729...34B}.   To expand on this wealth of information we must study more lenses.  The first step in doing this is to find more of these rare objects.

So far, less than a thousand lenses have been found in total across many heterogeneous data sets.
The Square Kilometer Array (SKA)\footnote{http://skatelescope.org/}, the Large Synoptic Survey Telescope (LSST)\footnote{https://www.lsst.org/} and the Euclid space telescope\footnote{https://www.euclid-ec.org/} are expected to increase the number of potential lenses by orders of magnitude \citep{2010MNRAS.405.2579O,collett_15,euclidSLWGwhitepaper,2015aska.confE..84M}.   For example it is estimated that there will be approximately 200,000 observable galaxy-galaxy lenses in the Euclid data set among tens of billions of potential objects.   These surveys will bring a new era for strong lensing where large relatively well defined samples of lenses will be possible.  It will also require handling much larger quantities of data than has been customary in this field.

Up to this point, the most widely used method for finding lenses in imaging surveys has been by visual inspection of candidates that have been selected on the basis of luminosity and/or colour.  This has been done in the radio \citep{2003MNRAS.341...13B} and in the visible with space and ground based data \citep{2008MNRAS.389.1311J,2008ApJS..176...19F,2010AandA...517A..25S,2014MNRAS.439.3392P}.  A related method pioneered by the SPACE WARPS project \citep{2016MNRAS.455.1171M,2016MNRAS.455.1191M,2015MNRAS.452..502G} has been to crowd source the visual inspection through an online platform.  These efforts have proved very fruitful, but they deal with orders of magnitude fewer candidates and lenses than will be necessary in the future.  Dealing with such large quantities of data will not be practical for any visual inspection approach.  In addition, the efficiency and detection bias of human inspection methods are difficult to rigorously quantify.

Spectroscopic searches for galaxy scale lenses have also been done by looking for high redshift stellar lines in the spectra of lower redshift large galaxies.  Notably this was done in the Sloan Lens ACS (SLACS) survey, producing a relatively well defined and pure sample of Einstein ring lenses \citep{2006MNRAS.369.1521W,2006ApJ...638..703B,2012ApJ...744...41B,2015MNRAS.449.3441S}.  
New spectrographs such as the Dark Energy Spectroscopic Instrument  (DESI) \citep{2016arXiv161100036D} and Subaru Prime Focus Spectrograph (PFS) \citep{2016SPIE.9908E..1MT} have the potential to greatly expand spectroscopic lens searches.
However, spectroscopy is telescope-time consuming and for the foreseeable future we are not likely to have spectroscopic surveys that cover anywhere near the number of objects as the planned imaging surveys.

Some automated algorithms have been developed in the past to detect lenses by their morphology in images.  These have been designed to detect arc-like 
features \citep{2006astro.ph..6757A,2007AandA...472..341S, 2017AandA...597A.135B} and rings 
\citep{2014ApJ...785..144G,2014AandA...566A..63J}.  They have been applied to survey data and found of order 200 lenses \citep{2007AandA...461..813C,2012ApJ...749...38M,2016AandA...592A..75P}.  

\cite{2009ApJ...694..924M} pioneered an automated technique for finding Einstein ring type lenses that uses a lens modelling code to fit a model to all candidates and picks out the ones that fit the model well \citep[see also][]{2017arXiv170401585S}.   This approach has the attractive feature that it distinguishes lenses from non-lenses by their similarity to what we expect a lens to look like and priors can be put on the model parameters that are physically motivated, unlike the next category of finders below.  Challenges arise in making the modelling fast and automatic enough to handle large 
data sets while allowing it to be flexible enough to find unusual lens configurations.  The YattaLens entrant to this challenge was of this type.

More recently, machine learning techniques that have become widely used in the fields of computer image processing and artificial intelligence have been applied to this  and other problems in astronomy; in particular, artificial neural networks (ANNs), support vector machines (SVM), and logistic regression.  SVMs and some logistic regression methods belong to the family of reproducing kernel Hilbert Space methods. They learn from a training set how to classify objects using features given by predefined kernel functions.  ANNs , and a popular variant convolutional neural networks (CNNs),  are even more flexible in learning directly from a training set which features are the most important for distinguishing categories of objects.  These have been used widely for such tasks as handwriting and facial recognition.  In astronomy, these families of algorithms are beginning to be used for categorising galaxy morphologies \citep{2015MNRAS.450.1441D}, photometric redshifts \citep{2017MNRAS.465.1959C,2016PASP..128j4502S,2017NewA...51..169S}, supernova classification \citep{2016ApJS..225...31L} and the lens finding problem \citep{2017arXiv170207675P,2017MNRAS.471..167J,2017MNRAS.465.4325O,2017AandA...597A.135B,hartley2017support}.

Given the future of this field, with large amounts of data coming soon and many new ideas emerging, it is timely to stage a series of challenges to stimulate new work, determine what can realistically be done in lens finding and get a better idea of the strengths and weaknesses of different methods.  The long term goal is to get a set of algorithms that can handle Euclid, LSST or SKA data sets and produce high purity and high completeness lens samples with well defined efficiency or selection.  We anticipate further challenges in the future in which the realism of the data simulations will become progressively better.  Here we have chosen to concentrate on galaxy/small group scale lenses where the background source is a galaxy because we feel that this is where the most progress can be made and the scientific return is the highest, but QSO lens and cluster/group lens challenges may follow.

The paper is organised as follows.  The form of the challenge and its rules are described in the next section.  The methods used to simulate mock images of galaxies and gravitational lenses are described in Section~\ref{sec:simulation}.  In Section~\ref{sec:entries}, each of the methods that were used to solve the challenge are briefly described.  We discuss the metrics used to evaluate entries in Section~\ref{sec:figure_of_merit}.  The performance of each of the methods is presented in Section~\ref{sec:performance}.  Finally,  in Section~\ref{sec:conclusion}, we conclude with a discussion of what was learned and how methods can be improved in the future.

\section{The Challenge}
\label{sec:challenge}

The challenge was in fact two separate challenges that could be entered independently.  One was designed to mimic a single band of a future imaging data set from a satellite survey such as Euclid.  The other was designed to mimic ground based data with multiple bands, roughly modeled on the Kilo-Degree Survey (KiDS)\footnote{http://kids.strw.leidenuniv.nl/} \citep{2013ExA....35...25D}. In neither case were the simulated images meant to precisely mock these surveys, but the surveys were used as guides to set noise levels, pixel sizes, sensitivities, and other parameters.

In each case, a training set of 20,000 images in each band was provided for download at any time along with a key giving some properties of the object including whether it was a gravitational lens.  Each image was $101\times101$ pixels.  These specifications were not of particular significance except that the image size would encompass almost all galaxy-galaxy lenses and that the number of images (including ones with and without noise, lens and source which were needed for later analysis) was not too large.
The participants were free to download these sets and train their algorithms on them.  To enter the contest, the participants needed to register with a team name at which point they would be given a unique key and the address of a test data set.  These data sets contained 100,000 candidates.  In the case of the multi-band ground-based set this was 400,000 images. The participants had 48 hours to upload a classification of all candidates consisting of a score between 0 and 1, 0 signifying the lowest confidence that it is a lens and 1 signifying the highest.  This ranking could have been a simple binary (0 or 1) classification  or it could have been a continuous range of numbers representing the probability of being a lens or it could have been a finite number of confidence levels.  
The challenge was opened on November 25, 2016 and closed on February 5, 2017.

\section{The simulations}
\label{sec:simulation}

Creating the mock images started with a cosmological N-body simulation, in this case the Millennium simulation \citep{2009MNRAS.398.1150B}.   A catalogue of dark matter halos and galaxies  within a light-cone was constructed within the Millennium Observatory project \citep{2013MNRAS.428..778O}.   The challenge sets were based on a 1.6 sq.deg. light cone extending out to redshift $z=6$ using all the simulation snapshots.  The halos were found with a friends-of-friends algorithm and characterised by a total mass, size and half mass radius.  They included subhalos of larger halos.  The halos where populated with galaxies based on their merger history using the semi-analytic model  (SAM) of \cite{2011MNRAS.413..101G}.

The halo catalogue was read into the GLAMER lensing code \citep{2014MNRAS.445.1942M,2014MNRAS.445.1954P} to do all the ray-tracing.    Within this code a Navarro, Frenk \& White (NFW) \citep{1996ApJ...462..563N} profile is fit to the three parameters given above to represent the dark matter component of the lens.  The halos are projected onto a series of 20 lens planes  and the deflection angle at any point on each plane are calculated by summing the effects of all the halos with a hybrid tree method.  In this way the halos have the mass, concentration and clustering properties from the N-body simulation, but within each strong lens the mass resolution is not limited by the original simulation, but follows the analytic mass profile.   An additional mass component that will be discussed later is added to each halo to represent the stellar mass.

With GLAMER we identify and map out all the caustics within the light-cone for 33 source planes -- z=1 to 3 in intervals of 0.1 and 3 to 6 in intervals of 0.25.  We take every caustic that corresponded to a critical curve with an Einstein radius larger than 1.5 times the resolution of the final images.  The Einstein radius is estimated here and in all that follows as $R_{\rm ein}=\sqrt{A_{\rm ein}/\pi}$ where $A_{\rm ein}$ is the angular area within the critical curve.

 In the light cone there are many thousands of caustics for the higher source redshifts.  These lenses could be used as is, but we wanted to produce a much larger number with more randomness.   For each caustic we identify the lens plane with the highest convergence and identify all the halos within a three dimensional distance of 0.5 Mpc from the centre of the critical curve and on this and its neighbouring lens planes.    This collection of halos is then used as the lens and rotated to produce more random lenses.    It  contains all the sub-halos and nearby companion halos, but not the large scale structure surrounding it.

To model the background objects that are lensed we use sources from
the Hubble Ultra Deep Field (UDF) that have been decomposed into
shapelet functions to remove noise.  This is the same set of images as
used in \citet{2008AandA...482..403M,2010AandA...514A..93M} \citep[see
also][]{2019MNRAS.482.2823P}.   There are 9,350 such sources with redshifts and separate shapelet coefficients in 4 bands.   

To construct a mock lens, first a caustic on the highest redshift source plane is selected.  This is done in order of Einstein area, but all the critical curves are used more than once.  Since every lens with a caustic at a lower redshift will have a caustic at the highest redshift this is a selection from all of the caustics in the light-cone.   The lens is extracted as explained above and rotated randomly in three dimensions.  A source is selected at random from the shapelet catalogue subject to a magnitude limit in a reference band.   The redshift of the UDF source is used as the source redshift.  If the source is at a lower redshift than the lens or within $\Delta z = 0.4$ another random source is selected.

The furthest point in the caustic is found from its own centre and the source is placed randomly within 3 times this distance.  This is a somewhat arbitrary length designed to be a compromise between producing only very clear strong lenses, because all the sources are right in the centre of the caustic, and making the process inefficient because most of the sources are too far away from the caustic to produce clear lenses.  If the source positions were taken completely at random the fraction of clear lenses would be very low.

The visible galaxies associated with the lens must also be simulated. There  are too few bright galaxies in the UDF catalogue  to make enough mock lens galaxies for this purpose.  Instead, for most of the lenses, we used an analytic model for the surface brightness of these galaxies.  The Millennium Observatory provides parameters for the galaxies that inhabit the dark matter halos using the semi-analytic galaxy formation models of \citet{2011MNRAS.413..101G}.  The parameters used here were the total magnitude, the bulge-to-disc ratio, the disc scale height and the bulge effective radius.  The magnitude and bulge-to-disc ratio are a function of the pass band.  Each galaxy is given a random orientation and inclination angle between 0 and $80\degree$.  The disc is exponential with no vertical height which is why the inclination is limited to $80\degree$.  The bulge is represented by an elliptical S\'{e}rsic profile with an axis ratio randomly sampled between 0.5 and 1.  The S\'{e}rsic index, $n_s$, is given by 
\begin{align}
\log( n_s )= 0.4 \log\left[ {\rm max}\left(\frac{B}{T},0.03\right)\right] + 0.1 x 
\end{align}
where $\frac{B}{T}$ is the bulge to total flux ratio and $x$ is a uniform random number between -1 and 1.  This very approximately reproduces the observed correlation between these quantities \citep{2001AJ...121..820G}. 

In addition to the basic disc and bulge models we introduce some spiral arms.  The surface brightness of 
the discs are given by
\begin{align}
S(\theta,r) &= e^{-r/R_h} \left[ 1 + A \cos(N_a\theta + \phi_r ) \right]~, \\
\phi_r &= \alpha\log(2 r/ R_h) + \phi_d \nonumber
\end{align}
where $R_h$ is the scale height of the disc.  The phase angle of the arms, $\phi_d$, is chosen at random.  The parameters $A$, $\alpha$ and $N_a$ are chosen from distributions that are judged by eye to produce realistic 
galaxies.
The bulge is also perturbed from a perfect S\'{e}rsic profile by multiplying the surface brightness by
\begin{align}
1+\sum_{n=1}^6 a_n \cos\left( n \theta  + \phi_n \right)
\end{align}
where $\phi_n$ is a random phase.  The coefficients are picked randomly from between -0.002 and 0.002.   

These foreground galaxies are rotated in three dimensions with the halos of the lens each time a random lens is produced so that they remain in the same positions relative to the mass.  All the random parameters are also reassigned with every realisation of the lens.

These images of the foreground galaxies are not intended to reproduce the true population of galaxies, but only to be sufficiently irregular to make them difficult to fit to a simple analytic model that might make them unrealistically easy to distinguish from a foreground plus a lensed image.   As will be discussed later, more realistic models will be needed in the future and are a subject of current investigation.

To represent the mass of the galaxies we make a gridded map of the surface brightness at 3 times the resolution of the final image.  The surface brightness map is converted into a mass map within GLAMER by assuming a uniform mass-to-light ratio of 1.5 times solar in the reference band.  These mass maps are added to the NFW dark matter halos discussed before to make the total lens mass distribution.  The deflections caused by the mass maps are calculated by Fast Fourier Transform (FFT) and added to the halos'  deflections for the ray tracing. 

The code is able to produce any combination of foreground galaxies, lensed image and noise that is desired.  For the training set,  an image of the total lens with noise, an image of the foreground galaxies with noise and image of the lensed background source without noise were provided.
For the test sets only the final images were provided to participants although all the information was stored for analysing the challenge entries.

\subsection{Space-based}
\label{sec:sim-space-based}

The space-based datasets were meant to roughly mimic the data quality which is expected from observations by the Euclid telescope in the visible channel. To this end, the pixel size was set to 0.1 arcsec and a Gaussian PSF was applied with a FWHM of 0.18 arcsec. The Gaussian PSF is clearly a simplified model, but a realistic treatment of the PSF is outside the scope of this paper. The reference band for background and foreground galaxies was SDSS $i$, which is overlapping with the broader Euclid VIS band. The realisation of the mock images followed the same procedure described in \citet{2004PASP..116..750G} and \citet{2008AandA...482..403M}. As a result, the noise follows a Gaussian distribution with a realistic width and is uncorrelated between pixels. Characteristics of the instrument, filter and exposure times were taken from the Euclid Red Book \citep{2011arXiv1110.3193L}.
 
 In the challenge set the limiting magnitude for background sources was 28 in $i$.   60\% of the cases had no background source and were thus labelled as non-lenses.
 
\subsection{Ground-based}
\label{sec:sim-ground-based}

For the ground-based images four bands (SDSS $u,g,r$, and $i$) where simulated.  The reference band was $r$.  For the challenge set, 85\% of the images where made with purely simulated images as outlined above and the other 15\% used actual images taken from a preliminary sample of bright galaxies directly from the KiDS survey.  Lensed source images where added to these real images at the same rate as for the mock images,  in this case 50\%.  No attempt was made to match the halo masses to the observed galaxies in these cases.  These real images where added for more realism and so that, by comparing the results for real and mock images, we can evaluate how realistic our simulations are in this context.  There were about 160,000 of these stamps from KiDS.

The KiDS survey provided a representative PSF map in each band that was applied to all mock images.   The pixel size in this case was 0.2~arcsec. 
Weight maps for the KiDS images were also provided.  Some of these had masked regions from removed stars, cosmic rays, and bad pixels.  For the mock images the noise was simulated by adding normally distributed numbers with the variance given by the weight maps.  The weight maps were also randomly rotated and flipped.  This resulted in many of the images having large masked regions in them.

By chance one of the original KiDS images appears to have been a lens.  When an additional lensed source was added this made a double lens or "jackpot" lens \citep{2008ApJ...677.1046G}.

\begin{table*}
\centering
\begin{tabular}{rllll}
  \hline
 & Name & type & authors & section \\ 
  \hline
  1 & AstrOmatic & Space-Based & Bertin  & \ref{sec:AstrOmatic}  \\ 
  2 & GAHEC IRAP & Space-Based & Cabanac & \ref{sec:IRAP} \\ 
  3 & CAS Swinburne Melb & Ground-Based & Jacobs   & \ref{sec:CASSwinburne} \\ 
  4 & ALL-star & Ground-Based & Avestruz, N. Li \& Lightman  & \ref{sec:ALL} \\ 
  5 & Manchester1 & Space-Based & Jackson \& Tagore  & \ref{sec:manchester} \\ 
  6 & CMU-DeepLens-Resnet-Voting & Space-Based & Ma, Lanusse \& C. Li  & \ref{sec:CMU_DeepLens} \\ 
  7 & Manchester SVM & Ground-Based & Hartley \& Flamary  & \ref{sec:Gabor-SVM} \\ 
  8 & CMU-DeepLens-Resnet & Space-Based & Francois Lanusse, Ma, C. Li \& Ravanbakhsh  & \ref{sec:CMU_DeepLens} \\ 
  9 & CMU-DeepLens-Resnet-Voting & Ground-Based & Ma, Lanusse \& C. Li  & \ref{sec:CMU_DeepLens} \\ 
  10 & YattaLensLite & Space-Based &  Sonnenfeld  & \ref{sec:YettaLens} \\ 
  11 & NeuralNet2 & Space-Based &  Davies \& Serjeant & \ref{sec:NeuralNet2} \\ 
  12 & CAST & Ground-Based &  Roque De Bom, Valent\'{\i}n \&  Makler  & \ref{sec:CAST} \\ 
  13 & CMU-DeepLens-Resnet-ground3 & Ground-Based &  Lanusse, Ma, Ravanbakhsh \& C. Li  & \ref{sec:CMU_DeepLens} \\ 
  14 & GAMOCLASS & Space-Based &  Huertas-Company, Tuccillo, Velasco-Forero \& Decenci\`{e}re & \ref{sec:GAMOCLASS} \\ 
  15 & LASTRO EPFL (CNN) & Space-Based & Geiger, Sch\"{a}fer \& Kneib  & \ref{sec:LASTRO} \\ 
  16 & Manchester SVM & Space-Based &  Hartley  \& Flamary & \ref{sec:Gabor-SVM} \\ 
  17 & CMU-DeepLens-Resnet-aug & Space-Based & Ma,  Lanusse, Ravanbakhsh \& C. Li  & \ref{sec:CMU_DeepLens} \\ 
  18 & LASTRO EPFL & Ground-Based & Geiger, Sch\"{a}fer \& Kneib  & \ref{sec:LASTRO} \\ 
  19 & CAST & Space-Based & Bom, Valent\'{\i}n \& Makler  & \ref{sec:LASTRO} \\ 
  20 & AstrOmatic & Ground-Based & Bertin  & \ref{sec:AstrOmatic} \\ 
  21 & ALL-now & Space-Based & Avestruz, N. Li \& Lightman  & \ref{sec:ALL} \\ 
  22 & Manchester2 & Ground-Based & Jackson \& Tagore  & \ref{sec:manchester} \\ 
  23 & YattaLensLite & Ground-Based & Sonnenfeld  & \ref{sec:YettaLens} \\ 
  24 & Kapteyn Resnet & Space-Based & Petrillo, Tortora, Kleijn, Koopmans \& Vernardos & \ref{sec:KapteynResnet} \\ 
   \hline
\end{tabular}
\caption{Entries to the challenges.  Descriptions of the methods are in the sections listed on the right.}
\label{table:entries}
\end{table*}

\section{Lens finding methods}
\label{sec:entries}

There were 24 valid entries into the challenge which are listed in Table~\ref{table:entries}.  There were a variety of different methods used and participants came from a variety of different backgrounds, most were professional astronomers, but there were also entries from researchers outside of the field.  

The following sections contain short descriptions of the lens finding methods that were used in 
the challenge.  Each sub-section refers to a team which gave a separate entry.  We have grouped the 
methods into four categories according to the type of method used.   The Receiver Operating Characteristic (ROC) curve and the area under this curve are referred to in these sections.  The ROC is defined in Section~\ref{sec:figure_of_merit} where methods for evaluating the entries is discussed.  A reader unfamiliar with the ROC might want to refer to that section.

\subsection{Visual inspection}
\subsubsection{Manchester1/Manchester2 (Jackson,Tagore)}
\label{sec:manchester}

All images (a total of 100000) were examined for each of the space- and
ground-based datasets. This was done by two observers; AT (Amit Tagore) examined 30000
images in each case and NJ (Neal Jackson) examined 70000. Observation was carried out
over a 48-hour period, at the rate of 5000/hr (NJ) and 2500/hr (AT). 
The overall results, in terms of area under the ROC (see Section \ref{sec:figure_of_merit}) curves, were very similar for 
both observers. The space-based challenge produced areas of 0.800 and 
0.812 for NJ and AT respectively, and the ground-based challenge yielded 
0.891 and 0.884.

The Python scripts used for manual examination of multiple images are available
on GitHub\footnote{https://github.com/nealjackson/bigeye} and are described in more detail
in \citet{hartley2017support}. For one-colour data such as
the space-based training set, the images are individually colour-scaled using
square-root scaling. The bright limit of the colour-scale is determined
from the pixel values in a rectangle comprising the inner ninth of the
image area, with the limit being chosen as the $n$th centile of the pixel
values in this area. Values between $n=95$ and $n=98$ give optimum results,
judging by experiments on the training set. The number of images in each
grid was also optimised using the training set, with 16$\times$8 or 
8$\times$4 giving good results on one-colour data. For three-colour data,
such as the ground-based challenge data, the individual bands for each 
object are colour-scaled and then combined into an RGB image. In this case
8$\times$4 grids were used for examination, due to the generally lower 
resolution of the images. The script also allows the user to adjust the
colour-scale in real time when examining and marking images, and records
the image name corresponding to the image within which the cursor resides
at the time any key is pressed, together with the key.

Images were classified by both observers into 5 categories, ranging from
0 (no evidence of any lensed objects in the image) to 4 (certain lenses).
For both observers, the rate of false positives in the ``certain'' lenses
was between 0.1\% and 0.3\%. The exception was the ground-based imaging 
for one observer, where a 4.6\% rate resulted mainly from a
single decision to allow a false-positive ``double lens'' which occurred
repeatedly throughout the data at different orientations. The false-negative
rate among the class-0 identifications was similar for both observers, at
around 25\% for the space-based images and 20\% for the ground-based.

\subsection{Arc-Finders}

These methods seek to identify gravitationally lensed arcs and differentiate between them and 
other objects such as spiral arms and edge on spirals using their width, colour, curvature and other pre-selected criterion.

\subsubsection{ GAHEC IRAP (Cabanac)}
\label{sec:IRAP}

\begin{figure}
 \includegraphics[width=\columnwidth]{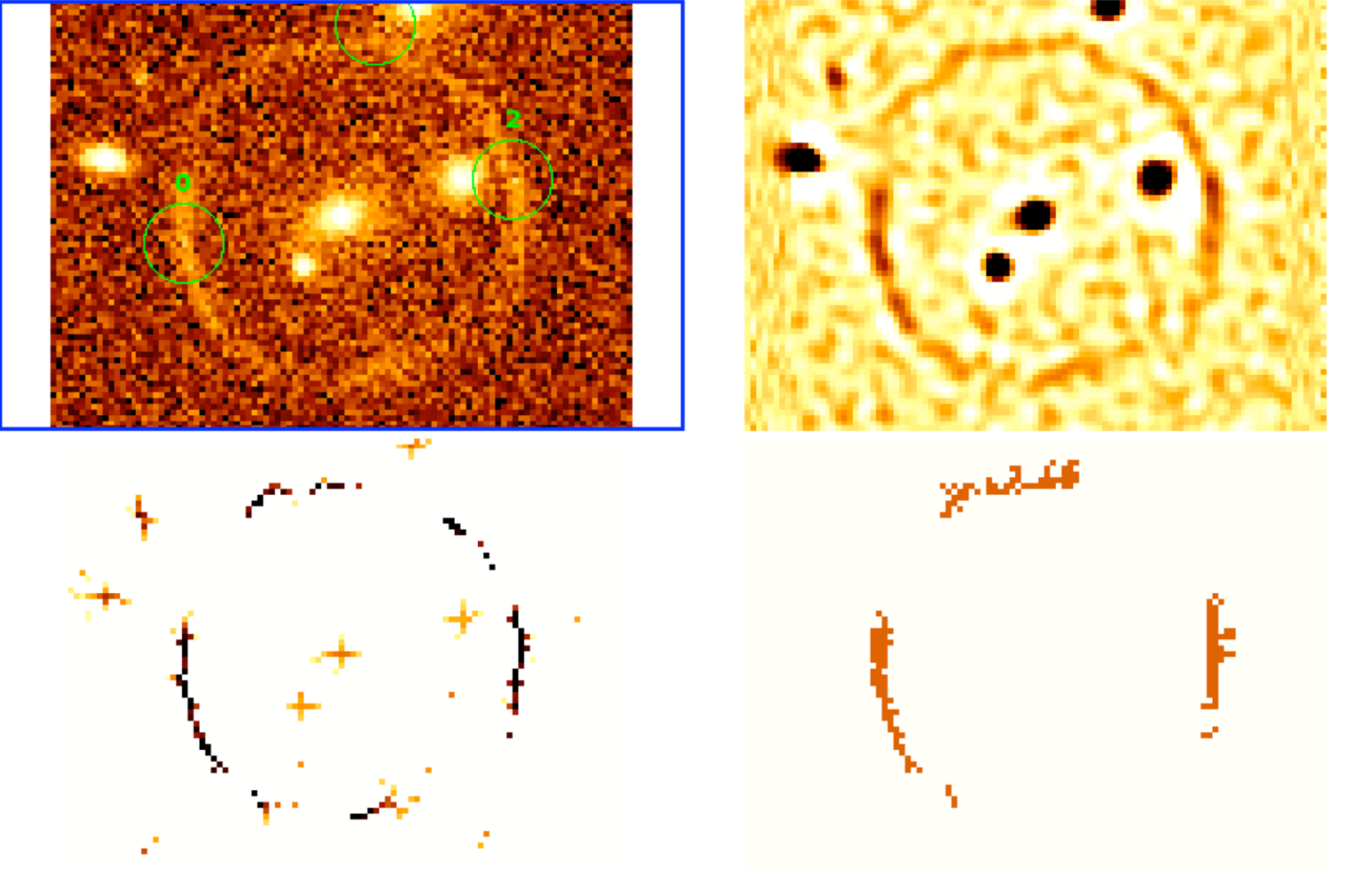}
 \caption{ (GAHEC IRAP) From top-left to bottom right, 1) a simulated arc extracted from the strong lensing challenge in which an tuned Arcfinder selects 3 candidates (green circles), 2) the smoothed image on which pixel wise elongation is computed, 3) the resulting elongated pixels after threshold, 4) the set of pixels selected for the computation of arc candidate properties. }
 \label{fig:Cabanac}
\end{figure}

Arcfinder \citep{2006astro.ph..6757A,2007AandA...461..813C,2012ApJ...749...38M} illustrated in Figure \ref{fig:Cabanac}, is a fast linear method that computes a pixel wise elongation parameter (ratio of first-order moments in a n-pix window oriented in proper reference frame) for all pixels of mexican-hat-smoothed FITS images. Arcfinder then extracts contiguous pixels above a given background and computes the candidate arc's length, width, area, radius of curvature and peak surface brightness. A final thresholding is set to maximize purity over completeness on a few typical arcs of the dataset.
For the current strong lensing challenge, arcfinder was tuned to detect long and narrow arcs, and was optimized on a subset of 1000 simulated images with a grid covering a range of elongation windows and arc areas.  A python wrapper allows users to change parameters in a flexible way and run the arcfinder C code from the linux  command line. Arcfinder took a couple of hours to run on the entire dataset with some overheads due to the dataset format. The code is publicly available at https://github.com/rcabanac/arcfinder.

\subsubsection{YattaLens Lite (Sonnenfeld)}
\label{sec:YettaLens}

YattaLensLite is a simpler version of the algorithm YattaLens \citep{2017arXiv170401585S}, modified to meet the time constraints of the challenge.
YattaLensLite subtracts a model surface brightness profile describing the lens galaxy from the $g$-band image, then runs SExtractor to detect tangentially elongated or ring-shaped objects, which are interpreted as lensed images.
In the ground-based challenge, the model lens surface brightness profile is obtained by taking a rescaled version of the $i$-band image.
The difference in colour between lens and source usually allows the lensed images to still be detectable after the lens subtraction process.
However, in order to avoid subtracting off the lensed images in systems with similar colours between lens and source, we radially truncate the model lens surface brightness.
The model lens light is truncated at the smallest radius between the position where the surface brightness is comparable to the sky background level, or the position of a positive radial gradient in surface brightness, if detected.

In the space-based challenge, it is not possible to separate lens and source based on colour, because only data in one band is provided. The lens light model then is produced by taking a centrally-inverted image and then using the same truncation prescription used with ground-based data. The central inversion step is taken to reduce the chances of subtracting flux from lensed images, which are in general not centrally symmetric as opposed to typical lens galaxies.

In the full version of YattaLens, a lens modelilng step is performed to improve the purity of the sample. However, such a procedure is too time consuming and was not performed in this challenge.

\subsection{Machine Learning Methods that use pre-selected features}

These are methods that classify the objects by making linear or nonlinear boundaries in a feature space.  The features are properties of the image and are typically chosen by the user with a combination of knowledge, intuition, and trial-and-error.  Using the training set, the optimal boundaries are found according to a criterion that depends on the method.  The machine learns how to use the features best for distinguishing between lenses and non-lenses.

\subsubsection{Manchester-SVM (Hartley, Flamary)}
\label{sec:Gabor-SVM}

A Support Vector Machine (SVM) is a supervised machine learning method which uses labelled training data to determine a classification model (see e.g., \citet{vapnik79estimation}, \citet{Cortes1995} and \citet{Burges1998}). A preprocessing stage first extracts a set of useful features from input samples, before projecting each sample as a vector into a high-, possibly infinite-dimensional space. The model then separates classes of data by maximising the margin between a defining hyperplane and a set of so-called support-vectors at the inner edge of each class. The process of classification is computationally inexpensive since the optimisation depends only on the dot products of the support vector subset. Feature extraction, however,  requires both an extensive exploration of the feature space during the development of a model, and potentially intensive computer resources in order to transform the original samples. Our full method is described in detail in \citet{hartley2017support} and was developed using the Python scikit-learn and scikit-image packages \citep{scikit-learn,scikit-image}.

During our development of an SVM classifier for lens finding, feature extraction initially involved the decomposition of each image into a set of objects, using SExtractor \citep{1996AandAS..117..393B} and GALFIT \citep{2002AJ....124..266P} to recover and subtract objects iteratively. This method had previously been used in a static algorithm approach which assigned points according to the morphological properties of each image \citep[see][]{2014AandA...566A..63J}. Lensed-like objects displaying, for example, greater ellipticity and tangential elongation were awarded more points. Since the SVM operates in a fixed dimensional space,  properties of individual objects were collapsed into a fixed set describing the mean and variance of morphological properties of all the objects within an image. After training an SVM using these features we recorded a modest separation of lens and non-lens classes.

An alternative approach was to design a set of Gabor filters to be applied to each sample. The Gabor kernel is described by a sinusoidal function multiplied by a Gaussian envelope. We discard the imaginary part of the function to leave, in two-dimensional space:
\begin{equation}
G_c[i,j]=Be^{-\frac{(i^2+j^2)}{2\sigma^2}} \mathrm{cos}\left[\frac{2\pi}{\lambda} (i\, \mathrm{cos} \, \theta + j\, \mathrm{sin} \,\theta)\right],
\end{equation}
where harmonic wavelength $\lambda$, Gaussian spread  $\sigma$ and orientation $\theta$ define the operation performed on each point $i,j$ in an image. Such a kernel is a popular image processing choice for edge detection and texture classification \citep[e.g.][]{Springer-verlag97computationalmodels,Feichtinger98a} and is thought to mimic some image processing functions of the mammalian brain \citep{Jones1233}.

\begin{figure}
  \centering
      \includegraphics[width=1\columnwidth]{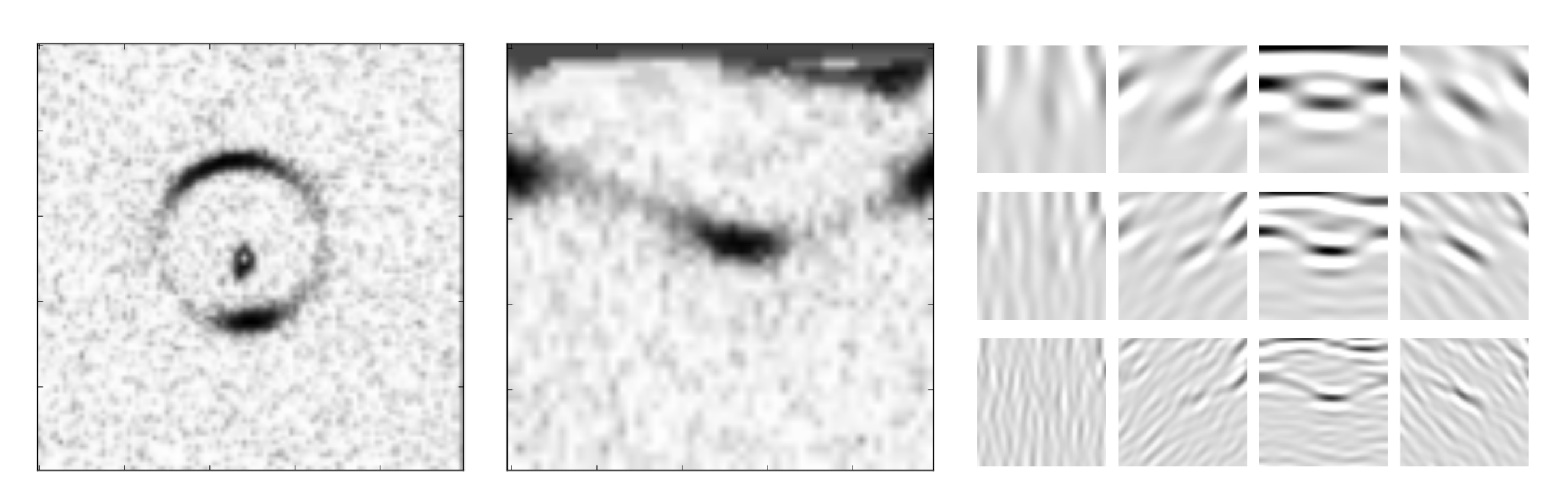} 
  \caption{Example of our feature extraction procedure used to transform a ring. The image on the right shows the response of a set of Gabor filters after convolution with a polar transformed image of an Einstein ring. The strongest response is seen in the orientation perpendicular to the radial direction and at the frequency most closely matching that of the ring.}
 \label{gaborring}
\end{figure}

Our final feature extraction procedure first applied a polar transform to each image in order to exploit the edge detection of the Gabor filter, picking out tangential components typical of galaxy-galaxy lensing. Each image was then convolved with several Gabor filters of varying frequency and rotation (see Fig.~\ref{gaborring}). Stability selection methods were used to investigate the classification performance using different combinations of filters. The responses for each Gabor filter when applied to each image were measured by calculating statistical moments for each filtered image. These moments formed  our final input data on which the SVM could be trained and applied. We used brute-force optimisation methods to select a non-linear SVM containing radial basis function (RBF) kernel and tuned a small set of regularisation hyperparameters to achieve good generalisation performance. During training and testing, our final and best scores achieved when testing on the  training data were an area under the ROC curve of 0.88 for the space set and 0.95 for the ground set. Classification was performed using a modest desktop PC.

\subsubsection{ALL (Avestruz, Li, Lightman)}
\label{sec:ALL}

The ALL team methodology is detailed in \citet{avestruz_etal17}. The
pipeline was originally developed to automatically classify strong
lenses in mock HST and LSST
data generated with code described in \citet{li_etal16} and
\citet{collett_15}.  We apply exactly the same steps for the single-band
data for Euclid, but modify the feature extraction step for the
four-band KIDS data.  We summarize the steps below.

Tools from {\em Scikit-learn} \citep{pedregosa_etal12} are used and
some minimal image preprocessing is  performed. First, we replace masked pixels
with the average of surrounding pixels, then enhance contrast in the
image by taking the normalized log of pixel values.  The next step 
consists of a feature extraction stage, where our feature
vector is a {\em histogram of oriented gradients} (HOG)
\citep{dalalandtriggs_05} that quantifies edges in the image.  HOG has
three main parameters that determine the binning and resolution of
edges captured by the features.  The result is a one dimensional
feature vector corresponding to the magnitude of oriented gradients
across the image.  With the KIDS data, we extract a feature vector for
each of the four bands and concatenate the vectors to create a final
feature vector for each object that we use to train a model
classifier.

We use {\em Logistic Regression} (LR) 
to train a classifier model.  LR requires a parameter search over the
regression coefficient, $C_{LogReg}$.  The parameters from both the
feature extrator, HOG, and the linear classifier, LR, contain
parameters that we optimize for peak model performance.  We use {\em
  GridSearchCV} from {\em Scikit-learn} to select cross-validated
parameters for HOG parameters and a subset of $C_{LogReg}$ values with
20\% of the test images provided.  We then run a finer parameter
search over $C_{LogReg}$, splitting the test images into 80\% training
and and 20\% test to avoid overfitting the data.  We use the best
parameters to then train the entire dataset for the final model
classifier that we used to evaluate the competition data.

\subsection{Convolutional Neural Networks}

Since Convolutional Neural Networks (CNNs) are central to many of the methods that will be described later, here we provide a brief general description of them.  A CNN \citep{Fukushima1980,Lecun1998} is a multi-layer feed-forward neural network model, which is particularly well-suited for processing natural images. With the very recent advances of the \textit{Deep Learning} framework \citep{Lecun2015}, models based on CNN architectures have reached or even surpassed human accuracy in image classification tasks \citep{He2015a}.

The fundamental building block of a CNN is the \textit{convolutional layer}. This element applies a set of convolution filters on an input image to produce a series of so-called \textit{feature maps}. The coefficients of these filters are free parameters that are learned by the model. The notation 
$n \times n - n_c$ will signify a convolutional layer using filters of size $n \times n$ pixels and outputting $n_c$ feature maps.  In typical architectures, the size of these convolution filters is kept small (i.e. 3x3 or 5x5 pixels) to limit the complexity of the model. 

Similarly to conventional fully-connected neural networks, convolution layers are typically followed  by an element-wise activation function, which allows for the modelling of complex functional forms  by introducing non-linearities in the model.  Classical choices  for  activation functions  include the sigmoid-shaped logistic  function (or just sigmoid) $f(x) = 1/ (1  + \exp(-x))$ or the hyperbolic tangent function $f(x) = \tanh(x)$. However, much of the success of Deep Learning is due to the  introduction of activation functions that do not saturate (become very close to one with very small derivatives), allowing for the  efficient training  of very deep architectures. The most common choice in modern deep learning models  is the simple ReLU activation (for rectified linear unit) \citep{Nair2010} defined as $f(x) = \max(x, 0 )$. A closely related  common alternative is the ELU activation (for Exponential Linear Unit) \citep{2015arXiv151107289C} defined as 
\begin{equation}
	f(x ) = \begin{cases}
    x & \text{if } x\geq 0\\
    e^{x } - 1,              & \text{otherwise}
\end{cases}\;,
\end{equation}
which often leads to better results in practice.

Because the filters used in convolution layers are  typically  just a few pixels in size, to capture features on larger scales, CNNs rely on a multi-resolution approach by interleaving convolutional layers with \textit{pooling layers}, which apply a downsampling operation to the feature maps. The most common downsampling schemes are the max-pooling and average pooling strategies, which downsample an input image by taking respectively the maximum or average values within a given region (e.g. 2x2 patches for a downsampling of factor 2).

A CNN architecture is therefore a stack of convolution layers and pooling layers, converting the input image into an increasing number of feature maps of progressively coarser resolution. The final feature maps can capture information on large scales and can reach a high-level of abstraction. To perform the classification itself from these feature maps, the CNN is typically topped by a fully-connected neural network outputting the class probability of the input image.

For a binary classification problem such as the one involved in strong lens detection, the training is performed by optimizing the weights of the model so that it  minimizes the \textit{binary cross-entropy} :
\begin{equation}
S =	- \sum_{n=1}^{N} y[n] \log \hat{y}[n] + (1 - y[n])\log(1 - \hat{y}[n]) \;,
\end{equation}
where $N$ is the number of training instances, $y \in \{0,1\}$ is the true class of the image and $\hat{y} \in [0,1]$ is the class probability predicted by the model. This optimization is usually performed by a Stochastic Gradient Descent (SGD) algorithm or its variants (e.g. ADAM \citep{Kingma_2014}, Adagrad \citep{duchi12adaptive}, RMSprop \citep{Tieleman12RMSProp}, or accelerated gradients \citep{nesterov83method}). SGD updates the model iteratively by taking small gradient steps over randomly selected subsamples of the training set (so called \textit{mini-batches}). All the CNN-based methods presented in this work rely on the ADAM optimisation algorithm, which also uses past gradients from previous iterations to adaptively estimate lower-order moments. Empirically it has been found that in many problems ADAM converges faster than SGD \citep{Ruder16overview}.

Neural networks often suffer from overfitting to the training set.  A common way to mitigate this is to use a regularisation scheme.  For example, the Dropout regularisation technique \citep{2012arXiv1207.0580H,JMLR:v15:srivastava14a}, were a certain percentage of the neurons and their connections are  randomly dropped  from the neural network. This regularisation techniques reduces overfitting  by preventing complex co-adaptations of neurons on training data. 

Training multi-layer neural networks with gradient descent based approaches can be very challenging. One of the main reasons behind this is the effect of \textit{vanishing gradients}: it has been empirically observed that in many multi-layer neural networks the gradients in higher-level (further from the image) layers often become too small to be effective in gradient descent based optimisation. Another difficulty is that the distribution of each layer's inputs changes during training as the parameters of the previous layers change. These issues make it difficult to find the best learning rates.

Batch normalisation layers \citep{batch_norm} are one way of addressing these challenges.  Let the activities of a given neuron in a mini-batch be denoted by $x_1,\ldots,x_m$. The batch normalisation layers i) calculate the empirical mean ($\mu = \frac{1}{m}\sum_{i=1}^m x_i $) and variance ($\sigma^2 = \frac{1}{m}\sum_{i=1}^m (x_i-\mu)^2 $) of the neural activities using the mini-batch data, ii) standardise the neuron activities to make them zero mean with unit variance, that is $\hat x_i = (x_i-\mu)/\sigma$, iii) linearly transform these activities with adjustable parameters $\beta, \gamma \in \mathbb{R}$: $y_i=\gamma\hat x_i +\beta$. Here $y_i$ denotes the output after applying the batch normalisation layer on the neuron with activities $x_i$. It has been empirically demonstrated that batch normalisation can often accelerate the training procedure and help mitigate the above described challenges.

There are many variations on these techniques and concepts some of which are represented in the following descriptions of particular methods.  

\subsubsection{AstrOmatic (Bertin)}
\label{sec:AstrOmatic}

The lens detector is based on a CNN, trained with the provided training datasets. The CNN is implemented in Python, using the TensorFlow framework\footnote{http://www.tensorflow.org/}. Both ground multichannel and space monochannel image classifiers have the exact same CNN architecture.

The network itself consists of three convolutional layers (11x11-32, 5x5-64 and 3x3-64), followed by two fully-connected layers ($256$ and $64$ neurons) and an output softmax layer. The first five layers use the ELU activation function, which in our tests led to significantly faster convergence compared to ReLU and even SoftPlus activation. Dropout regularization is applied to both convolutional and fully connected layers, with ``keep'' probabilities $p=2/3$ and $p=1/2$, respectively.

Prior to entering the first convolutional layer, input image data are rescaled and the dynamic-range compressed with the function $f(x) =
\mathrm{arcsinh} (10^{11} x)$, and bad pixels are simply set to 0.
Data augmentation (increasing the amount of training data by modifying and reusing it) is performed in the form of random up-down and left-right image flipping, plus $k\pi/2$ rotations, where $k$ is a random integer in the $[0,3]$ range. Additionally, a small rotation with random angle $\theta$ is applied, involving bicubic image resampling. the angle $\theta$ is drawn from a Gaussian distribution with mean $\mu=0$ and standard deviation $\sigma_{\theta}=5^{\circ}$. No attempt was made to generate and randomize bad pixel masks in the data augmentation process.


The CNN weights are initialized to random values using a truncated Gaussian distribution with mean $\mu=0$ and standard deviation $\sigma=0.05$. The network is trained on a Titan-X ``Pascal'' nVidia GPU using the ADAM gradient-based optimizer during 800 epochs, with an initial learning rate $\eta(t=0)=10^{-3}$ and a learning rate decay $\eta(t+1)/\eta(t)=0.99$, where $t$ is the epoch. Because of a lack of time, tests were limited to assessing the basic classification performance on a subset of the of 1,000 images/datacubes, using the 19,000 others for training.

\subsubsection{LASTRO EPFL (Geiger, Sch\"{a}fer)}
\label{sec:LASTRO}

We used a CNN \citep{Fukushima1980,Lecun1998} with a simple architecture of 12 layers (inspired by \citep{symmetry}), see table \ref{tab:architecture}.
To avoid the problem of the data flow distribution getting out of the comfort zone of the activation functions ("Internal Covariate Shift"), we used a mix of normalization propagation \citep{norm_prop} (without the constraint on the weights but a proper initialization) and batch normalization \citep{batch_norm} (slowly disabled over the iterations).
As activation function, we used a scaled and shifted ReLU, defined as
\begin{equation} \label{eq:relu}
    \frac{1}{\sqrt{\pi-1}} (\sqrt{2 \pi} \max(0,x) - 1),
\end{equation}
to satisfy the properties required by the normalization propagation.
Our batch normalization implementation computes the mean of the activation function $\bar\mu_i$ using the following equation 
\begin{equation} \label{eq:batchnorm}
    \bar\mu_i \longleftarrow (1-\eta) \; \bar\mu_{i-1} + \eta \; \mu_i(\text{batch}).
\end{equation}
$\bar\mu_i$ is computed using the mean value $\mu_i$ over the batch in combination with the previous mean $\bar\mu_{i-1}$ using an inertia value $\eta$ set to $1$ at the beginning and decaying with the iterations.
For the training, the 20,000 provided images were split into two sets, 17,000 for training and 3k for validation.
Each iteration of the gradient descent (more precisely ADAM \citep{adam}) minimizes the cross entropy, 
\begin{equation} \label{eq:xent}
    \left\{
    \begin{array}{ll}
        - \log(p)   & \text{if the image is a true lens} \\
        - \log(1-p) & \text{if the image is a nonlens}
    \end{array}
    \right.,
\end{equation}
where $p$ is the output of the neural network, computed over a batch of 30 images, 15 lenses and 15 nonlenses, picked from the training set.
The small batches with only 30 images were easier to handle computationally but added more noise to the gradient which we considered negligible due to there being only two classes to classify.

To augment the training set, each image of the batch is transformed with a random transformation of the dihedral group (rotations of 90 degrees and mirrors), its pixel values multiplied by a factor picked between $0.8$ and $1.2$ and shifted by a random value between $-0.1$ and $0.1$.
To prevent the overfitting, we used some dropout \citep{dropout} (with a keeping probability decreasing with the iterations).
The masked regions of the ground based images are handled by simply setting them to zero.
Each final prediction is made of the product of the predictions of the 8 transformations of the image by the dihedral group.
The architecture is implemented in Tensorflow\footnote{\url{http://tensorflow.org/}}.
Our code is accessible on github\footnote{\url{https://github.com/antigol/lensfinder-euclid}}. Additional details can be found in \citet{2017Schaefer}.

\begin{table*}
    \centering
    \begin{tabular}{|l|l|l|l|}
        \hline
        Layer type & shape & activation & \# parameters \\ \hline \hline
        
        \textbf{convolutional 4x4} & $101\!\times\! 101\!\times\!1/4 \to 98\!\times\!98\!\times\!16$ & rectifier & 256/1'024 + 16 \\ \hline
        \textbf{convolutional 3x3} & $98\times98\times16 \to 96\times96\times16$ & rectifier & 2'304 + 16 \\ \hline
        max pool /2 & $96\times96\times16 \to 48\times48\times16$ & - & - \\ \hline
        batch normalization & $48\times48\times16$ & - & 16 + 16 \\ \hline
        
        \textbf{convolutional 3x3} & $48\times48\times16 \to 46\times46\times32$ & rectifier & 4'608 + 32 \\ \hline
        \textbf{convolutional 3x3} & $46\times46\times32 \to 44\times44\times32$ & rectifier & 9'216 + 32 \\ \hline
        max pool /2 & $44\times44\times32 \to 22\times22\times32$ & - & - \\ \hline
        batch normalization & $22\times22\times32$ & - & 32 + 32 \\ \hline
        
        \textbf{convolutional 3x3} & $22\times22\times32 \to 20\times20\times64$ & rectifier & 18'432 + 64 \\ \hline
        \textbf{convolutional 3x3} & $20\times20\times64 \to 18\times18\times64$ & rectifier & 36'864 + 64 \\ \hline
        max pool /2 & $18\times18\times64 \to 9\times9\times64$ & - & - \\ \hline
        batch normalization & $9\times9\times64$ & - & 64 + 64 \\ \hline
        dropout & $9\times9\times64$ & - & - \\ \hline
        
        \textbf{convolutional 3x3} & $9\times9\times64 \to 7\times7\times128$ & rectifier & 73'728 + 128 \\ \hline
        dropout & $7\times7\times128$ & - & - \\ \hline
        \textbf{convolutional 3x3} & $7\times7\times128 \to 5\times5\times128$ & rectifier & 147'456 + 128 \\ \hline
        batch normalization & $5\times5\times128$ & - & 128 + 128 \\ \hline
        dropout & $5\times5\times128$ & - & - \\ \hline
        
        \textbf{fully-connected} & $5\times5\times128 \to 1024$ & rectifier & 3'276'800 + 1'024 \\ \hline
        dropout & $1024$ & - & - \\ \hline
        \textbf{fully-connected} & $1024 \to 1024$ & rectifier & 1'048'576 + 1'024 \\ \hline
        dropout & $1024$ & - & - \\ \hline
        \textbf{fully-connected} & $1024 \to 1024$ & rectifier & 1'048'576 + 1'024 \\ \hline
        batch normalization & $1024$ & - & 1'024 + 1'024 \\ \hline
        \textbf{fully-connected} & $1024 \to 1$ & sigmoid & 1'024 + 1 \\ \hline \hline
        Total & - & - & $\approx$ 5'674'000 \\ \hline
    \end{tabular}
    \caption{LASTRO EPFL architecture}
    \label{tab:architecture}
\end{table*}

\subsubsection{GAMOCLASS (Tuccillo, Huertas-Company, Velasco-Forero, Decenci\`ere)}
\label{sec:GAMOCLASS}

\begin{figure*}
  \centering
      \includegraphics[width=1\textwidth]{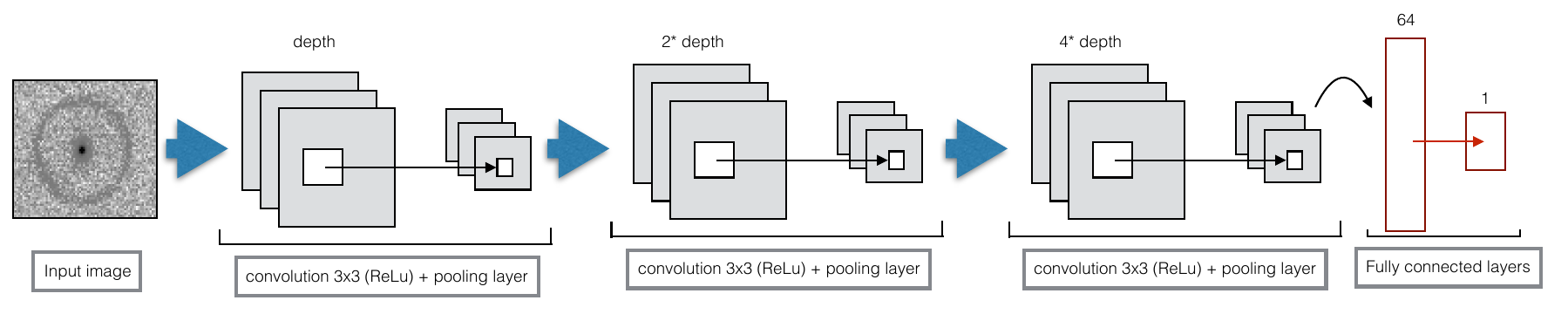} 
  \caption{GAMOCLASS schematic}
 \label{Fig1_gamoclass}
\end{figure*}

GAMOCLASS is a CNN based classifier.
We used the full training data set in the proportion of 4/5 for training and 1/5 for validation. The training images were labelled with 1 if showing strong lensing and 0 otherwise. Our CNN gives as output a probability [0,1] of the input image being a strongly lensed system.
The final architecture of our model is illustrated in Fig. \ref{Fig1_gamoclass}. The input image (101x101 pixels) is first processed by a 2D convolution layer with a 3x3 filter size, then subsampled by a 3x3 max pooling layer.  Another two identical units follow, with a growing dimensionality of the output space 
 in the convolution, for a total of 3 convolutional layers and 3 max
 pooling layers. Each of these convolutional layers is followed by a ReLU step. The output of these units is then processed through a single fully-connected layer follower by a dropout layer, and, finally, by a one-neuron fully connected layer with sigmoid activation functions. For the classification problem we used the binary cross-entropy cost function and found the weights  using ADAM \citep{Kingma_2014}) optimization method. The use of the ADAM optimizer improved the learning rate compared to tests with stochastic gradient descent (SGD).
In order to increase the size of the training set and make the model invariant to specific transformations, we perform these data augmentation steps:
1) we introduce random rotations of the image in the range [0, 180$\degree$], using a reflection fill mode to keep constant the size of the images;
2) the images are randomly shifted of 0.02 times the total width of the image;
3) the images are randomly flipped horizontally and vertically.

During the training we initialize the weights of our model with random normal values and we ``warm up" the training of the CNN for 25 epochs, using an exponential decay rate ($10^{-6}$) \citep{Huang_2016}  and then a staring learning rate of  0.001. Then the network was trained using an early stopping method, and for a maximum number of 300 epochs. The early stopping method is an effective method of preventing overfitting and consists in stopping the training if a monitored quantity does not improve for a fixed number (called \textit{patience}) of training epochs. The quantity that we monitored was the \textit{accuracy} of the classification of the validation sample. The best architecture was trained over 220 epochs with a parameter of patience equal to 20.
We implemented our code in the Keras framework \citep{Chollet_2015} on top of Theano \citep{Bastien_2012}.
Our architecture converges with a classification accuracy of 91\% on the validation sample. We further evaluated the performance of our classifier calculating the ROC (see Section \ref{sec:figure_of_merit})  curve of the classifier, i.e. the True Positive Rate (TPR) against the False Positive Rate.  We reached a TPR higher than the 90\% with a FPR < 8\%.

\subsubsection{CAST (Bom, Valent\'in, Makler)}
\label{sec:CAST}

The CBPF Arc Search Team (CAST) tested several arcfinding schemes with CNNs at their core.  
For both the space-based and ground-based samples we used a simple preprocessing phase to enhance the objects in the images to check to 
see if this improved the automated arc detection with the CNN. We chose a contrast adjustment with $0.1\%$ pixel saturation and apply a low pass band Wiener filter \citep{wiener1964extrapolation} to reduce the effect of the noise. 

We used a native CNN from Matlab\footnote{\texttt{https://www.mathworks.com/products/matlab.html}, \texttt{https://www.mathworks.com/help/nnet/convolutional-neural-networks.html}}, which has Convolutional 2D layers with 20 5x5 filters. This CNN can work either with one or three input images, representing greyscale and colour images.
We employed different strategies for the two samples available for the challenge, which involve combinations of the available bands running in one or more CNN, using or not the preprocessing, and combining the output with the aid of other machine learning methods.
In each case we used the simulations made available for the challenge both to train and to validate the results and we used the area under the ROC (see Section \ref{sec:figure_of_merit}) to determine which combination of methods gives the best result.    
We selected 
$90\%$ of the images, chosen randomly, for the training and $10\%$ to validate. We repeated the process $10$ times to avoid bias due to a specific choice of training/validation set and to define an uncertainty in our ROC.

For the space-based data set we tested only two configurations: i) using the CNN straightforwardly for classification and ii) 
with the preprocessing described above. 
We found that the results, accounting for the uncertainties,  
were clearly superior in terms of the area under the ROC with the preprocessing. Therefore, this is the configuration we used for the challenge entry. 

As mentioned above, the CNN we used can take 3 colour images as input.  To use the information on the $4$ available bands, we needed to either combine $2$ of the $4$ bands to end up with $3$ bands for a single RGB CNN (configuration I below) or we use multiple CNNs (configuration II to VI below). To combine the outputs of several CNNs we use a Support Vector Machine \citep[ SVM; see e.g. ][]{rebentrost2014quantum} also implemented in Matlab. The SVM is used to combine the outputs $p_i$ of the several CNNs (configurations II, III, IV and VI). Instead of using only $p_i$ as inputs to the SVM we also tested providing the SVM with image features obtained by the CNN (feature maps, configuration V) as inputs. In all cases we tested with and without the preprocessing

A more detailed description of each configuration tested is presented below:
\begin{enumerate}[label=\Roman*.]
\item Combination of bands $r$ and $i$ with the average between bands $u$ and $g$. Use one CNN for classification.
\item Creation of $1$ CNN for each band (total of $4$). The $4$ outputs are used as input to a SVM classifier which returns the final classification $p$. 
\item Combination into 4 different combinations of bands:  RGB $\rightarrow$ (\textit{u,g,r}), (\textit{u,g,i}), (\textit{u,r,i}) and (\textit{g,r,i}). One CNN for each combination of bands and then use of the output score as input to an SVM classifier.
%
\item  Average of bands in different combinations RGB $\rightarrow$ (\textit{ug,r,i}), (\textit{u,gr,i}) and (\textit{u,g,ri}). The outputs of these 3 CNNs are inputs to a SVM classifier.
\item Use of CNN-activations (CNN feature maps) as inputs to a SVM classifier, using same combinations of bands of III. The output of each CNN is used as input to a SVM classifier.
\item Use of wiener filter and contrast adjustment on each band, then using the resulting images in the same architecture as in (III). 
\end{enumerate}

\begin{figure}
 \includegraphics[width=\columnwidth]{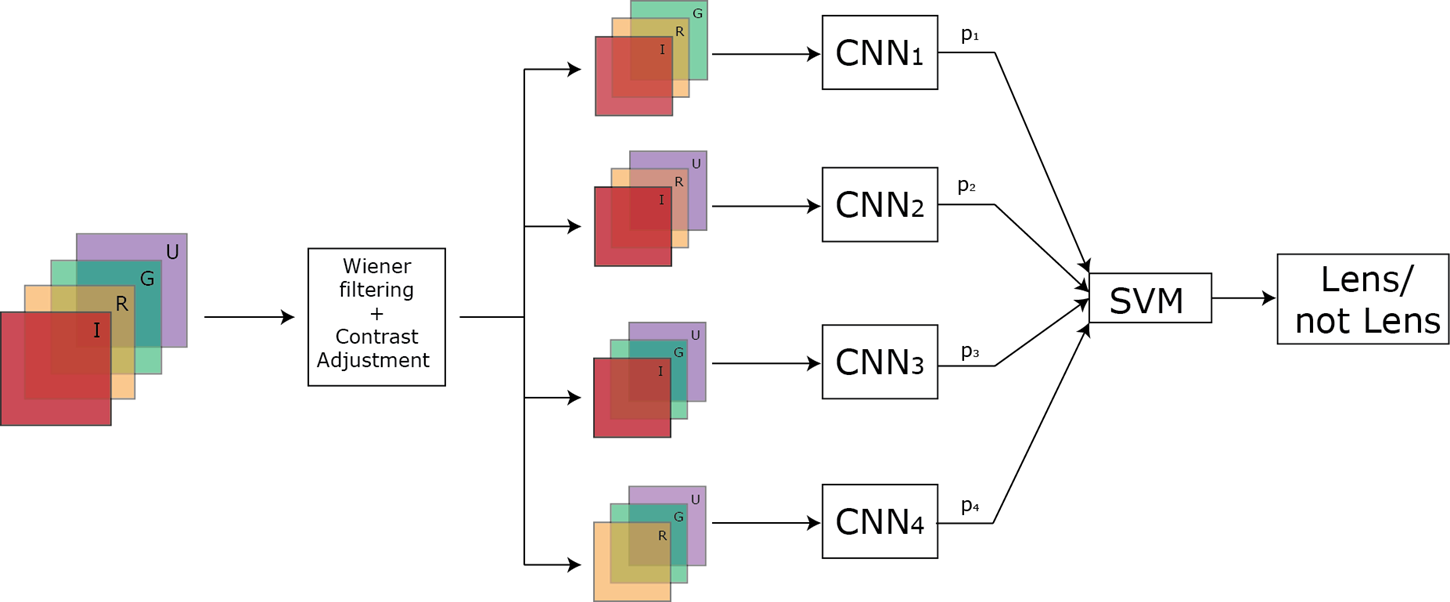}
 \caption{CAST Lens finder pipeline for the ground based sample. Illustration of the chosen architecture for the CAST search in the case of ground-based simulated images.}
 \label{colorCASTCNN}
\end{figure}

For the ground based cases, the three configurations with highest area under ROC were III, IV and VI. Although the areas are very similar between IV and VI the last one is superior in the low fake positives end. Thus, for the Strong Lensing Challenge in the ground base sample, we used configuration VI. This final scheme is illustrated in figure \ref{colorCASTCNN}.

The area under ROC, in both space based configurations were, in general, smaller than in the multi-band case, which suggests how the CNNs are sensitive to colour information to find strong lensing. Particularly,  ground base configuration II used one CNN per band and has the similar area under ROC as our best single band configuration. 



\subsubsection{CMU DeepLens (Lanusse, Ma, Li, Ravanbakhsh)}
\label{sec:CMU_DeepLens}


\texttt{CMU DeepLens} is based on a residual network (or \textit{resnet}) architecture \citep{He2015a}, a modern variant of CNNs which can reach much greater depths (over 1000 layers) while still gaining accuracy. We provide a short overview of our model below but a full description of our architecture can be found in \citet{Lanusse2017}.

Much like conventional CNNs, resnets are based on convolutional and pooling layers. However, resnet differ from CNNs by the introduction of so-called shortcut connections bypassing blocks of several convolutional layers. As a result, instead of learning the full mapping from their input to their output these residual blocks only have to  learn the difference to the identity transformation. In practice, this difference allows residual networks to be trained even for very deep models. For a more thorough description of this architecture, we refer the interested reader to Section~2.3 of \citet{Lanusse2017}.

Our baseline model is composed of a first 7x7-32 convolutional layer which can accommodate either single-band or multi-band images. The rest of the model is composed of 5 successive blocks, each block being made of 3 resnet units (specifically, pre-activated bottleneck residual units \citep{He2016}). At each block, the signal is downsampled by a factor 2 and the number of feature maps is in turn multiplied by 2. The model is topped by an average-pooling layer followed by a single fully-connected sigmoid layer with a single output. Apart from the final layer, we use the ELU (Exponential Linear Unit) activation throughout. The weights of the model are initialized using random normal values, following the strategy advocated in \citet{He2015a}. 

Training was performed using the ADAM optimizer with mini-batches of size 128 over 120 epochs, with an initial learning rate of $\alpha=0.001$, subsequently decreased by a factor 10 every 40 epochs. This multi-step training procedure is important to progressively refine the model parameters and achieve our final accuracy.

We adopt a minimal pre-processing strategy for the input images, removing the mean image and normalising by the noise standard deviation $\sigma$ in each band, this statistic being evaluated over the whole training set. In addition, we clip extreme values above $250 \,\sigma$ to limit the dynamic range of the input. Bad-pixels are simply set to 0 after this pre-processing step.

Given the relatively small training set preventing overfitting is an important consideration. In our final model, we combine several data augmentation strategies: random rotations (in the range $[-90, \ 90^\circ]$), random mirroring along both axes, and random resizing (by a small factor in the range $[0.9, \ 1]$).

The architecture presented above is the one that lead to our best results in both branches of the challenge, i.e. \texttt{CMU-DeepLens-Resnet} for space-based and \texttt{CMU-DeepLens-Resnet-ground3} for ground-based. We also submitted results for two variants of this baseline model, named \texttt{-aug} and \texttt{-Voting}. 
The first variant introduced several data-augmentation schemes, including the ones mentioned above and the addition of Gaussian noise to the input images. We found however that the introduction of noise was not necessary as the other methods were enough to prevent overfitting. 
The second variant was used to explore a voting strategy between three different models. These models differed by the type of residual blocks (bottleneck vs wide) and by their handling of missing pixels (setting to 0 or to the median value of the image). The predictions of the best 2 out of 3 models were then averaged to produce the final classification probability.


Our model is implemented using the \texttt{Theano}\footnote{\url{http://deeplearning.net/software/theano/}} and \texttt{Lasagne}\footnote{\url{https://github.com/Lasagne/Lasagne}} libraries. On an Nvidia Titan Xp GPU, our full training procedure requires approximately 6 hours on the ground-based challenge, but classification of the whole testing set is performed in a couple of minutes. Finally, in the interest of reproducible research, our code is made publicly available on GitHub\footnote{\url{https://github.com/McWilliamsCenter} }.
This repository also contains a notebook detailing how to reproduce our challenge submission.

\subsubsection{Kapteyn Resnet (Petrillo, Tortora, Vernardos, Kleijn, Koopmans) }
\label{sec:KapteynResnet}

Our lens-finder is based on a CNN, following the strategy adopted recently in \citep{2017arXiv170207675P}. 
We decide to treat the problem as a three-class classification problem where the classes are \textit{non-lenses}, \textit{clear lenses} and \textit{dubious lenses}. We define the \textit{dubious lenses} as the lenses with lensing features with less than 160 pixels and the \textit{clear lenses} those with more than 160 pixels belonging to the lensed source. This choice is motivated by the fact that specializing the network in recognizing different classes could lead to a more robust classification. In addition, in a hypothetical application of the method to real data from a survey, this could be a way to select the most blatant lenses.  


The CNN is implemented in Python 2.7 using the open-source libraries \textsc{Lasagne}\footnote{\href{http://github.com/Lasagne/Lasagne/}{\tt http://github.com/Lasagne/Lasagne/}} 
and \textsc{Theano}
\footnote{\href{http://deeplearning.net/software/theano/}{\tt http://deeplearning.net/software/theano/}} \citep{theano}.  
The training of the CNN is executed on a GeForce GTX 760 in parallel with the data augmentation performed on the CPU using the \textsc{scikit-image}\footnote{\href{http://scikit-image.org/}{\tt http://scikit-image.org/}} package \citep{van2014scikit}. 

We used the CNN architecture called Resnet described in \citep{he2015deep} with three stacks of residual blocks of 5 layers. 
The output layer is composed by three units. Each unit gives as an output a number between 0 and 1 that represents, respectively, the probability of being a \textit{non-lens}, a \textit{dubious lens}, a \textit{certain lens}.  We then collapsed one of the classes into another to give a binary classification : 0 when a source is classified as a \textit{non-lens} and a 1 when is classified as a \textit{clear lens} or as a \textit{dubious lens}. This choice did not allow for building  a continuous ROC (see Section \ref{sec:figure_of_merit})  curve but only a binary one.
The final submission was produced by averaging the values of the predictions from three CNNs with the same architecture.

The training image files were preprocessed with the software \textsc{STIFF}\footnote{\href{http://www.astromatic.net/software/stiff}{\tt http://www.astromatic.net/software/stiff}} which automatically converts the fits files to grey-scale TIFF images operating a non-linear intensity transformations to enhance the low-brightness features of the image.   
Due to memory limitations we down-sampled the images to 84 by 84 pixels. 
We augmented the training images in the following way: i) random rotation of 90, 180 or 270 degrees; ii) random shift in both $x$ and $y$ direction between -2 and +2 pixels; iii) $50\%$ probability of horizontally flipping the image. 
Finally, the image border is cropped in order to have 80 by 80 pixel input images.

The network is trained by minimizing the categorical cross-entropy loss function
\begin{equation}
L = -\sum_j t_{j}\log p_{j}
\label{EQloss}
\end{equation}
where the $t_j$ and $p_j$ are respectively the label and the prediction for the class $j$. 
The minimization is done via mini-batch stochastic gradient descent with ADAM updates \citep{Kingma_2014}. We used a batch size of 56 and performed  46000 gradient updates. We started with a learning rate of $4 \times 10^{-4}$, decrease it to $4 \times 10^{-5}$ after 35000 updates and to $4 \times 10^{-6}$ after 43000 updates.
The weights of each filter are initialized from a random normal distribution with variance ${2/n}$ where $n$ is the number of inputs of the unit  and a mean of zero \citep{He2015a}.  
We use L2-norm regularization with $\lambda= 9 \times 10^{-3}$.

\subsubsection{NeuralNet2 (Davies, Serjeant) }
\label{sec:NeuralNet2}

Our lens finder included wavelet prefiltering. The image was convolved with the Mallat wavelet with a kernel size of 4 in both the horizontal and vertical directions, then combined and compared to the original image to make the input image;  input image $= \sqrt[]{H^{2}+V^{2}}$. This prefiltering was performed to emphasise the edges in the images. It was found to improve the results compared to the CNN without this pre-filter. The CNN had 2 convolution layers each containing  $3\times3 - 32$ filters, incorporating dropout and max-pooling, and then 3 dense fully-connected layers to classify each image. The network was trained on 18000 of the 20000 training images; training took 15 epochs and was completed once the validation loss was minimised. The training was validated on the remaining 2000 images. Validation loss was calculated using binary cross entropy
\begin{equation}
\mathcal{L} = \sum_{i=1}^{n} \big[ y_{i} \log(p_{i}) + (1  -  y_{i}) \log(1  -  p_{i}) \big]
\end{equation} 
where $\mathcal{L}$ is the loss function, $n$ is the number of inputs, $y_{i}$ is the true value of the $i^{th}$ input , and $p_{i}$ is the predicted value for the $i^{th}$ input from the network. A perfect loss of $0$ was generated once every predicted value matched the true value for every input. The network was made and trained in Python 2.7 using the open-source libraries \textsc{theano} and \textsc{keras}\footnote{\href{https://github.com/keras-team/keras}{\tt https://github.com/keras-team/keras}}. A more developed version of our lens finder will appear in Davies, Bromley and Serjeant (in preparation).

\subsubsection{CAS Swinburne (Jacobs)}
\label{sec:CASSwinburne}

Our model is a CNN-based classifier. The architecture of our network was simple, similar to that of AlexNet \citep{krizhevsky_imagenet_2012}, with three 2D convolutional layers (with kernel sizes 11, 5, and 3), and two fully-connected layers of 1024 neurons each. The activation function after each convolutional layer was a ReLU. After each convolution we employed a 3x3 max pooling layer. To avoid over-fitting, we included a dropout layer of 0.5 after each of the two fully-connected layers. We implemented our network using the Keras python framework \citep{Chollet_2015} and Theano \citep{Bastien_2012}.

The training set was augmented with three rotations, and 20\% of the images were reserved for validation. The training set consisted of 4-band FITS files of simulated lenses and non-lenses. We imported the training set into HDF5 database files. The data was normalised on import, such that the mean value of the data cube, across all bands, is zero and the standard deviation is one, i.e. \(X' = (X - \mu)/\sigma\); the dynamic range was not altered. We also include batch normalisation step after the first convolution, which normalises the outputs of this layer to the same range (\(\mu = 0\), \(\sigma = 1\)). This has been shown empirically to aid in more rapid convergence of the training process.

The training process using a categorical cross-entropy loss function, and a stochastic gradient descent optimizer with an initial learning rate of 0.01, learning rate decay of \(10^{-6}\), and Nesterov momentum \citep{nesterov83method} of 0.9. Training converged (validation loss stopped decreasing) after approximately 30 epochs.

We note that experiments indicated that training on 4-band FITS data, as opposed to RGB images produced from the fits files, resulted in improved validation accuracy, of order a few percent. 

\section{Results}
\label{sec:results}

In this section we summarize the analysis of the submissions.  In Section~\ref{sec:figure_of_merit} we discuss how to judge a classifier in this particular case and define some metrics of success.  The results for all the submissions are given in Section~\ref{sec:performance}.

\subsection{Figures of merit}
\label{sec:figure_of_merit}

In deriving a good figure of merit for evaluating lens finding algorithms one needs to take into account the particular nature of this problem.
The traditional method for evaluating a classification algorithm is with the receiver operating characteristic curve, or  ROC  curve.  This is a plot of the true positive rate (TPR) versus the false positive rate (FPR).  In this case these are defined as
\begin{align}
{\rm TPR} &= \frac{\textrm{ number of true lenses classified as lenses}}{\textrm{ total number of true lenses}} \\
{\rm FPR} &= \frac{\textrm{number of non-lenses classified as lenses}}{\textrm{ total number of non-lenses}}
\end{align}
The classifier generally gives a probability of a candidate being a lens, $p$, in which case a threshold is set and everything with $p$ greater is classified as a lens and everything smaller is classified as not a lens.  The TPR and FPR are then plotted as a curve parametrised by this threshold.  At $p=1$ all of the cases are classified as non-lenses and so TPR=FPR=0 and at $p=0$ all of the cases are classified as lenses so TPR=FPR=1.  These points are always added to the ROC curve.  If the classifier made random guesses then the ratio of lenses to non-lenses would be the same as the ratio of the number of cases classified as lens to the number of cases classified as non-lenses and so TPR=FPR.  The better a classifier is the smaller the FPR and the larger the TPR so the further away from this diagonal line it will be.  When a classifier provides only a binary classification or a discrete ranking, the ROC connects the endpoints to the discrete points found by using each rank as a threshold.

A common figure of merit for a classifier is the area under the ROC ({\bf AUROC}).  This evaluates the overall ability of a classifier to distinguish between cases.  This was the criterion on which the challenge participants were told to optimise.  However, in the case of gravitational lensing this is not the only thing, and not the most important thing, to consider.  Gravitational lenses are rare events, but to improve the discrimination and training of the classifiers the fraction of lenses in test and training sets are boosted to something around half.  In these circumstances it is important to consider the absolute number of cases that will be misclassified when the fraction of true cases is closer to what is expected in the data.

If the rates of false positives and false negatives remain the same in real data  the contamination of the sample will be
\begin{align}
\frac{\rm FP}{\rm TP} \simeq \frac{\rm FPR}{\rm TPR} \left(\frac{\textrm{number of non-lenses in sample}}{\textrm{number of lenses in sample}} \right).
\end{align}
Since only about one in a thousand objects will be a lens (perhaps somewhat more depending on pre-selection) the contamination will be high unless the FPR is much less than the TPR.  For this reason we consider some additional figures of merit.

The {\bf TPR$_0$} will be defined as the highest TPR reached, as a function of $p$ threshold, before a single false positive occurs in the test set of 100,000 cases.  This is the point were the ROC meets the FPR = 0 axis.  This quantity highly penalizes classifiers with discrete ranking which often get TPR$_0$ = 0 because their highest classification level is not conservative enough to eliminate all false positives.  We also define {\bf TPR$_{10}$} which is the TPR at the point were less than ten false positives are made.  If the TP rate is boosted from the FPR by a factor of 1,000 in a realistic data set this would correspond to about a 10\% contamination.

In addition to these considerations, the performance of a classifier is a function of many characteristics of the lens system.  It might be that one classifier is good at finding systems with large Einstein radii and incomplete arcs, but not as good at finding small complete Einstein rings that are blended with the light of the lens galaxy.  Also a lens may have a source that is too faint to be detected by any algorithm or is too far from the lens to be very distorted, but will be classified as a lens in the test dataset.    We do not impose a definitive arc/ring magnification, brightness or surface brightness limit for a system to be considered a lens because we 
want to include these ``barely lensed" objects to test the limits of the classifiers.  As we will see, if one restricts the objectives to detecting only lensed images with surface brightness above some threshold, for example, the "best" algorithm might change and the TPR will change.  For this reason we plot the AUROC, TPR$_0$ and TPR$_{10}$ as a function of several variables for all the entries.  This is done by removing all the lenses that do not exceed the threshold and then recalculating  these quantities, while the number of non-lenses remains the same.

\subsection{Performance of the methods}
\label{sec:performance}

\begin{table*}
\centering
\begin{tabular}{llrrrl}
  \hline
  Name & type & AUROC & TPR$_0$ & TPR$_{10}$ & short description \\ 
  \hline
 CMU-DeepLens-Resnet-ground3 & Ground-Based & 0.98 & 0.09 & 0.45 & CNN \\ 
  CMU-DeepLens-Resnet-Voting & Ground-Based & 0.98 & 0.02 & 0.10 & CNN \\ 
  LASTRO EPFL & Ground-Based & 0.97 & 0.07 & 0.11 & CNN \\ 
  CAS Swinburne Melb & Ground-Based & 0.96 & 0.02 & 0.08 & CNN \\ 
  AstrOmatic & Ground-Based & 0.96 & 0.00 & 0.01 & CNN \\ 
  Manchester SVM & Ground-Based & 0.93 & 0.22 & 0.35 & SVM / Gabor \\ 
  Manchester2 & Ground-Based & 0.89 & 0.00 & 0.01 & Human Inspection \\ 
   ALL-star & Ground-Based & 0.84 & 0.01 & 0.02 & edges/gradiants and Logistic Reg. \\ 
   CAST & Ground-Based & 0.83 & 0.00 & 0.00 & CNN / SVM \\ 
   YattaLensLite & Ground-Based & 0.82 & 0.00 & 0.00 & SExtractor \\ 
   LASTRO EPFL & Space-Based & 0.93 & 0.00 & 0.08 & CNN \\ 
  CMU-DeepLens-Resnet & Space-Based & 0.92 & 0.22 & 0.29 & CNN \\ 
   GAMOCLASS & Space-Based & 0.92 & 0.07 & 0.36 & CNN \\ 
  CMU-DeepLens-Resnet-Voting & Space-Based & 0.91 & 0.00 & 0.01 & CNN \\ 
  AstrOmatic & Space-Based & 0.91 & 0.00 & 0.01 & CNN \\ 
   CMU-DeepLens-Resnet-aug & Space-Based & 0.91 & 0.00 & 0.00 & CNN \\ 
   Kapteyn Resnet & Space-Based & 0.82 & 0.00 & 0.00 & CNN \\ 
   CAST & Space-Based & 0.81 & 0.07 & 0.12 & CNN \\ 
  Manchester1 & Space-Based & 0.81 & 0.01 & 0.17 & Human Inspection \\ 
   Manchester SVM & Space-Based & 0.81 & 0.03 & 0.08 & SVM / Gabor \\ 
   NeuralNet2 & Space-Based & 0.76 & 0.00 & 0.00 & CNN / wavelets \\ 
   YattaLensLite & Space-Based & 0.76 & 0.00 & 0.00 & Arcs / SExtractor \\ 
   All-now & Space-Based & 0.73 & 0.05 & 0.07 & edges/gradiants and Logistic Reg. \\ 
  GAHEC IRAP & Space-Based & 0.66 & 0.00 & 0.01 & arc finder \\ 
   \hline
\end{tabular}
\caption{The AUROC, TPR$_0$ and TPR$_{10}$ for the entries in order of AUROC.}
\label{table:AUROC}
\end{table*}

Table~\ref{table:AUROC} shows the AUROC, TPR$_0$ and TPR$_{10}$ for the entries in order of AUROC and dataset type.  It can be seen that CMU-DeepLens-Resnet-ground3 had the best AUROC for the ground-based set and LASTRO EPFL the best for the space-based set.  The order is different if TPR$_0$ is used to rank the entries as seen in table~\ref{table:TPR0}.  Here Manchester SVM and 
CMU-DeepLens-Resnet get the best scores.

\begin{table*}
\centering
\begin{tabular}{llrrrl}
  \hline
  Name & type & AUROC & TPR$_0$ & TPR$_{10}$ & short description \\ 
  \hline
 Manchester SVM & Ground-Based & 0.93 & 0.22 & 0.35 & SVM / Gabor \\ 
  CMU-DeepLens-Resnet-ground3 & Ground-Based & 0.98 & 0.09 & 0.45 & CNN \\ 
  LASTRO EPFL & Ground-Based & 0.97 & 0.07 & 0.11 & CNN \\ 
   CMU-DeepLens-Resnet-Voting & Ground-Based & 0.98 & 0.02 & 0.10 & CNN \\ 
   CAS Swinburne Melb & Ground-Based & 0.96 & 0.02 & 0.08 & CNN \\ 
   ALL-star & Ground-Based & 0.84 & 0.01 & 0.02 & edges/gradiants and Logistic Reg. \\ 
   Manchester2 & Ground-Based & 0.89 & 0.00 & 0.01 & Human Inspection \\ 
   YattaLensLite & Ground-Based & 0.82 & 0.00 & 0.00 & SExtractor \\ 
   CAST & Ground-Based & 0.83 & 0.00 & 0.00 & CNN / SVM \\ 
   AstrOmatic & Ground-Based & 0.96 & 0.00 & 0.01 & CNN \\ 
   CMU-DeepLens-Resnet & Space-Based & 0.92 & 0.22 & 0.29 & CNN \\ 
   GAMOCLASS & Space-Based & 0.92 & 0.07 & 0.36 & CNN \\ 
   CAST & Space-Based & 0.81 & 0.07 & 0.12 & CNN \\ 
   All-now & Space-Based & 0.73 & 0.05 & 0.07 & edges/gradiants and Logistic Reg. \\ 
   Manchester SVM & Space-Based & 0.80 & 0.03 & 0.07 & SVM / Gabor \\ 
   Manchester1 & Space-Based & 0.81 & 0.01 & 0.17 & Human Inspection \\ 
   LASTRO EPFL & Space-Based & 0.93 & 0.00 & 0.08 & CNN \\ 
   GAHEC IRAP & Space-Based & 0.66 & 0.00 & 0.01 & arc finder \\ 
   AstrOmatic & Space-Based & 0.91 & 0.00 & 0.01 & CNN \\ 
   Kapteyn Resnet& Space-Based & 0.82 & 0.00 & 0.00 & CNN \\ 
   CMU-DeepLens-Resnet-aug & Space-Based & 0.91 & 0.00 & 0.00 & CNN \\ 
   CMU-DeepLens-Resnet-Voting & Space-Based & 0.91 & 0.00 & 0.01 & CNN \\ 
   NeuralNet2 & Space-Based & 0.76 & 0.00 & 0.00 & CNN / wavelets \\ 
   YattaLensLite & Space-Based & 0.76 & 0.00 & 0.00 & Arcs / SExtractor \\ 
   \hline
\end{tabular}
\caption{The AUROC, TPR$_0$ and TPR$_{10}$ for the entries in order of TPR$_0$.  }
\label{table:TPR0}
\end{table*}

\begin{figure*}
 \includegraphics[width=2\columnwidth]{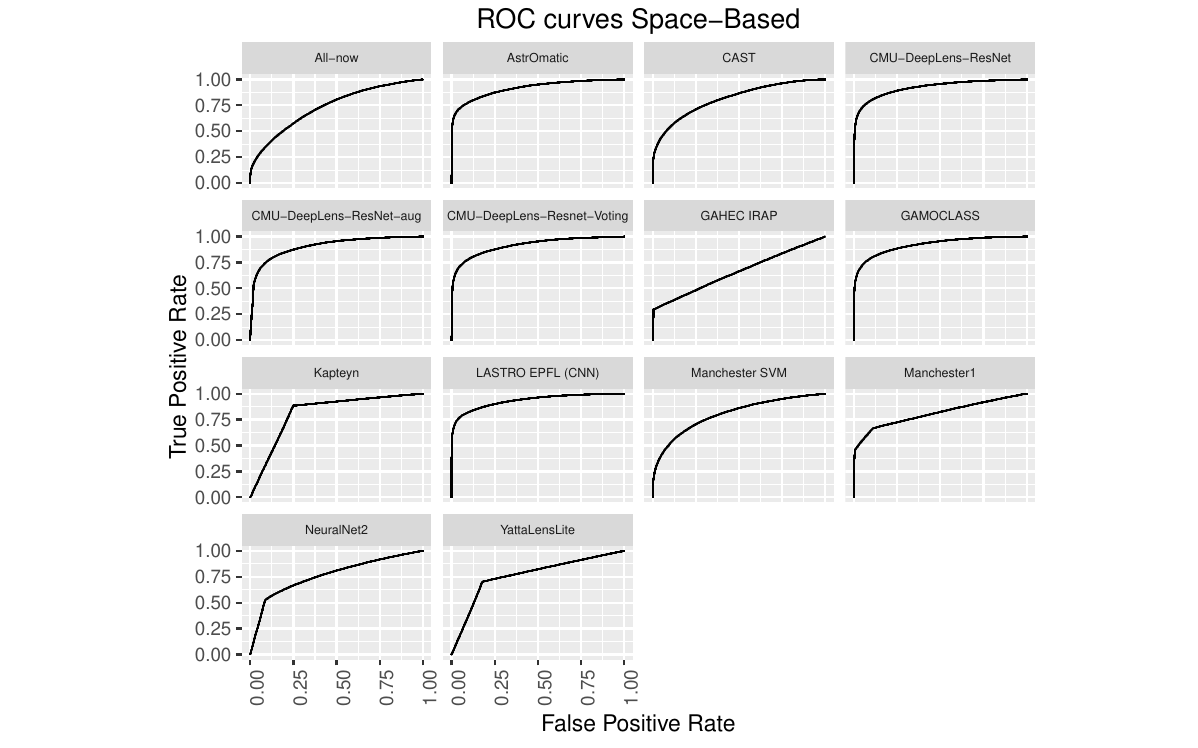}
 \caption{ROC curves for the space-based entries.}
 \label{fig:roc_space}
\end{figure*}

\begin{figure*}
 \includegraphics[width=2\columnwidth]{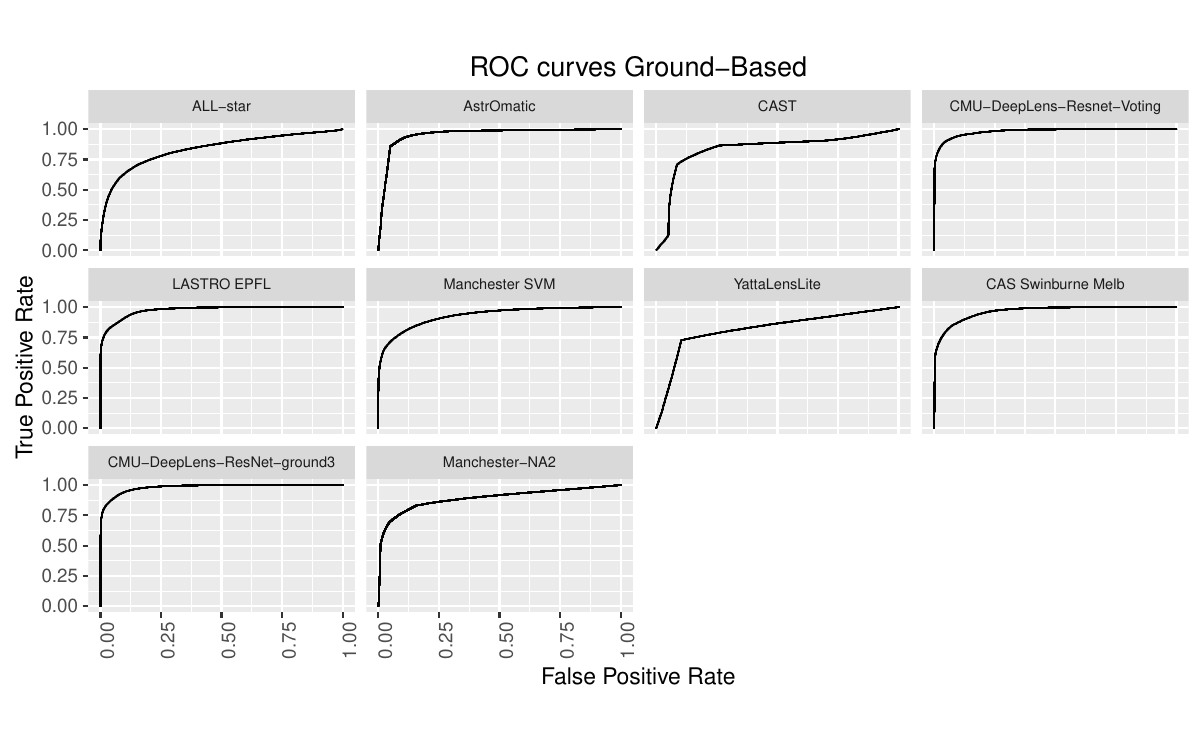}
 \caption{ROC curves for the ground-based entries.  Notice that these are generally better than in figure~\ref{fig:einstein_space} indicating that colour information is an important discriminant. }
 \label{fig:roc_ground}
\end{figure*}

\begin{figure*}
 \includegraphics[width=2\columnwidth]{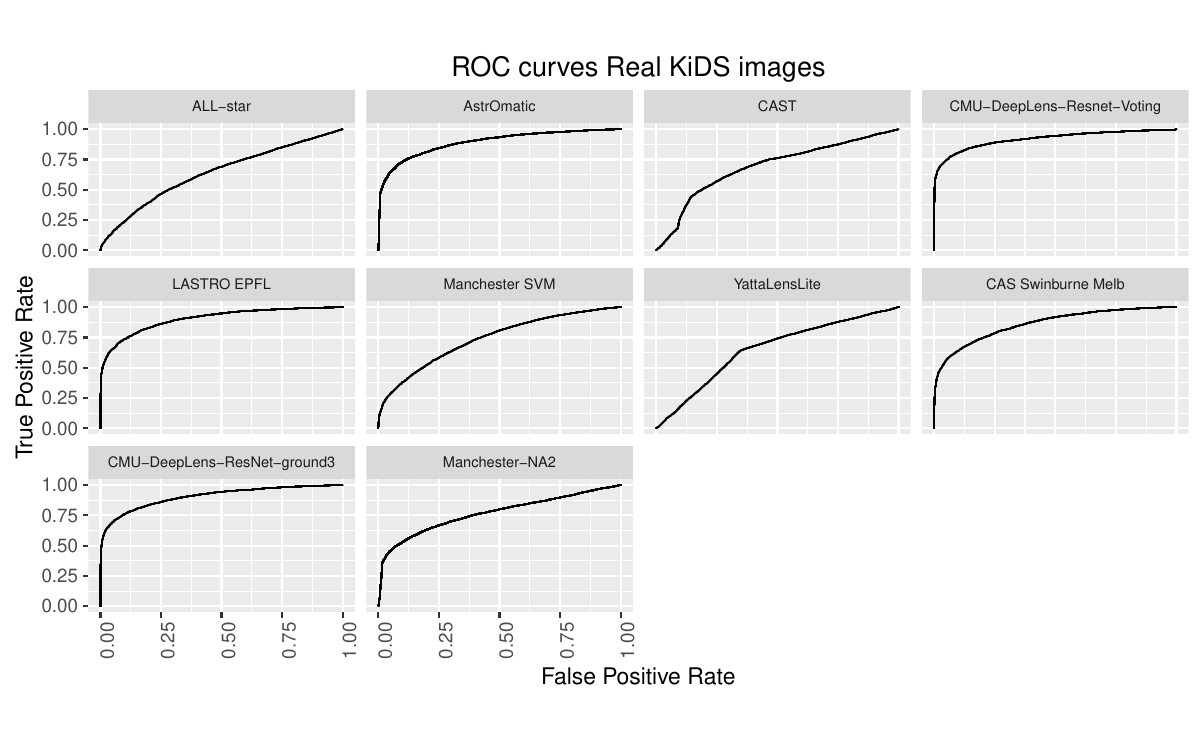}
 \caption{ROC curves for the ground-based entries including only the cases with authentic images taken from the KiDS survey.  It can be seen that in all cases these are lower than in figure~\ref{fig:roc_ground}.}
 \label{fig:roc_kids}
\end{figure*}

Figures~\ref{fig:roc_space} and \ref{fig:roc_ground} show the ROC curves for all the entries.  
We note that ROC curves for the ground-based challenge (figure~\ref{fig:roc_ground}) are uniformly better than those for the space-based challenge (figure~\ref{fig:roc_space}).  This is because of the importance of colour information in discriminating lensed arcs from pieces of the foreground lens galaxy.

In addition, figure~\ref{fig:roc_kids} shows the ROC curves for only the ground-based images where an actual KiDS image was used (see Section~\ref{sec:sim-ground-based}).  It can be seen that the classifiers do uniformly less well on this subset.  This indicates that the simulated galaxy images are different from the real ones and that the classifiers are able to distinguish fake foreground galaxies from lenses more easily than from real galaxies.  Some methods are more affected by this than others, but none seem to be immune, not even the human classification. This is perhaps not unexpected, but does show that the simulated lenses need to be improved before the raw numbers can be directly used to evaluate the performance of a classifier on real data.

\begin{figure*}
 \includegraphics[width=2\columnwidth]{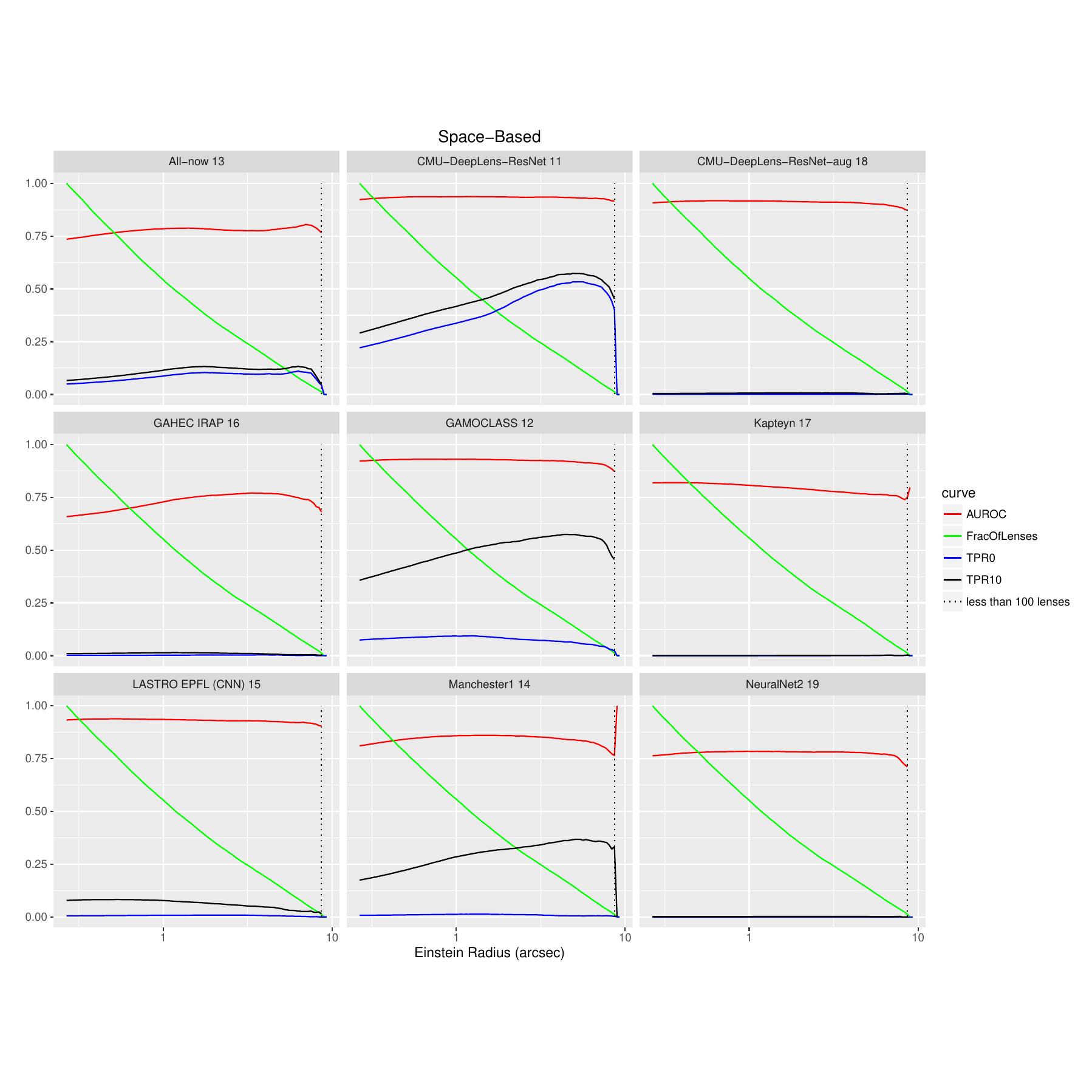}
 \caption{The AUROC, TPR$_0$, TPR$_{10}$ and the fraction of lenses in the test sample after discarding the lenses with Einstein radii larger than the number indicated on the x-axis. The vertical dotted lines indicate where no more than 100 lenses in the test sample had larger Einstein radii.  Beyond this point one should be suspicious of small number statistics.  }
 \label{fig:einstein_space}
\end{figure*}

\begin{figure*}
 \includegraphics[width=2\columnwidth]{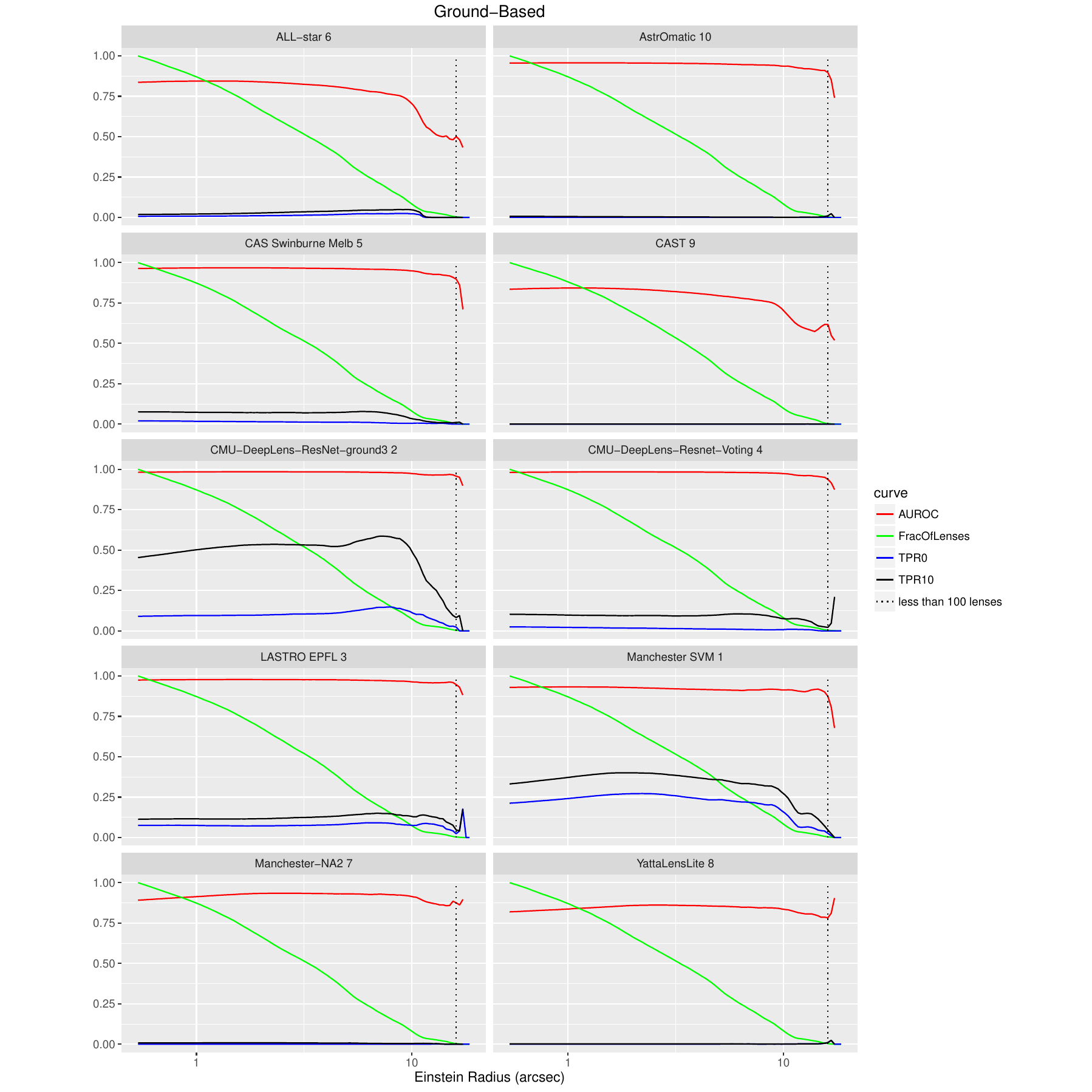}
 \caption{Same as figure~\ref{fig:einstein_space}, but for ground-based entries.  The AUROC, TPR$_0$, TPR$_{10}$ and the fraction of lenses in the test sample after discarding the lenses with Einstein radii larger than the number indicated on the x-axis. The vertical dotted lines indicate where no more than 100 lenses in the test sample had larger Einstein radii.  }
 \label{fig:einstein_ground}
\end{figure*}

\begin{figure*}
 \includegraphics[width=2\columnwidth]{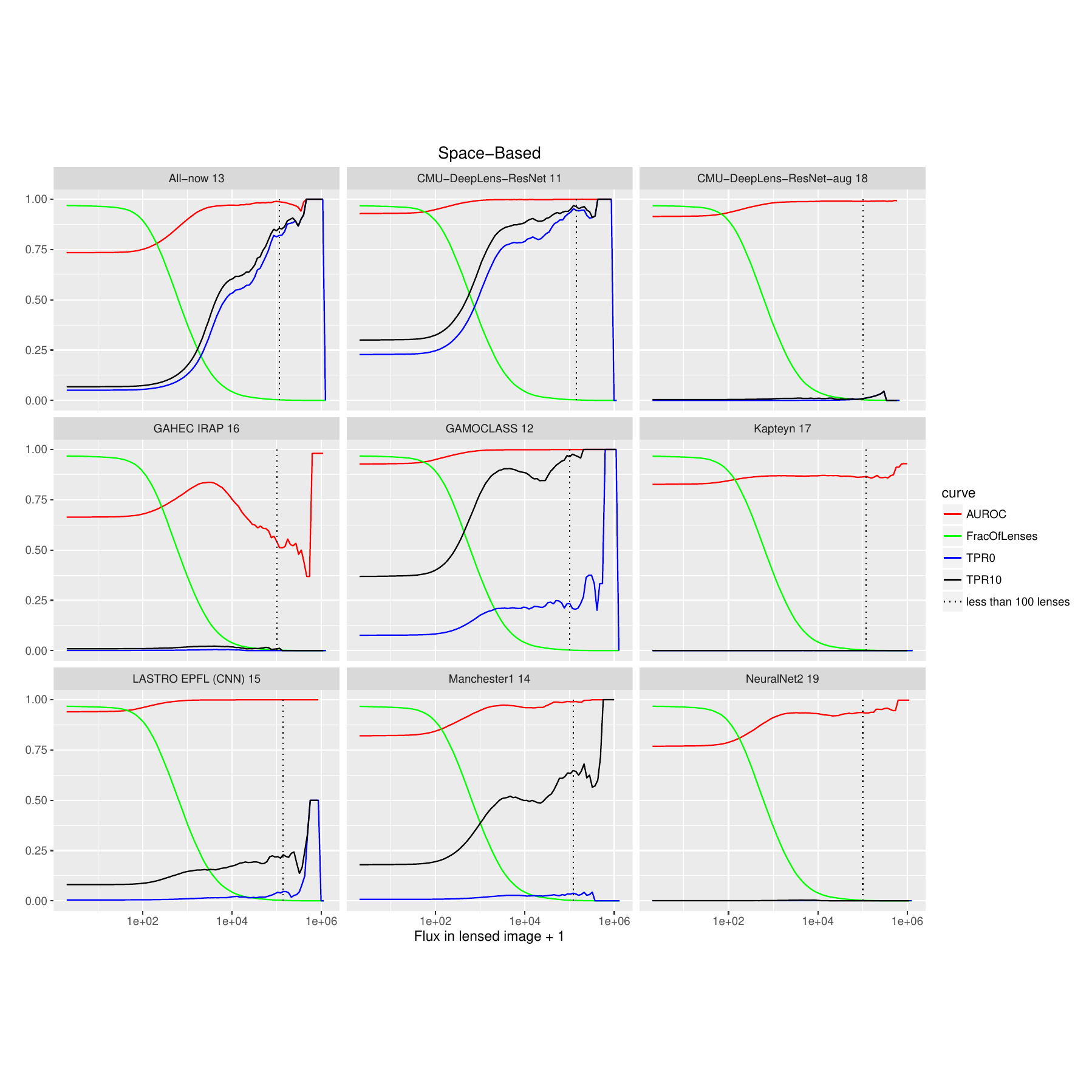}
 \caption{ The AUROC, TPR$_0$, TPR$_{10}$ and the fraction of lenses in the space-based test sample after discarding the lenses with fluxes within the pixels that are above 1 $\sigma$ in the lensed source as indicated on the x-axis. The vertical dotted lines indicate where no more than 100 lenses in the test sample had larger Einstein radii. }
 \label{fig:flux_space}
\end{figure*}

\begin{figure*}
 \includegraphics[width=2\columnwidth]{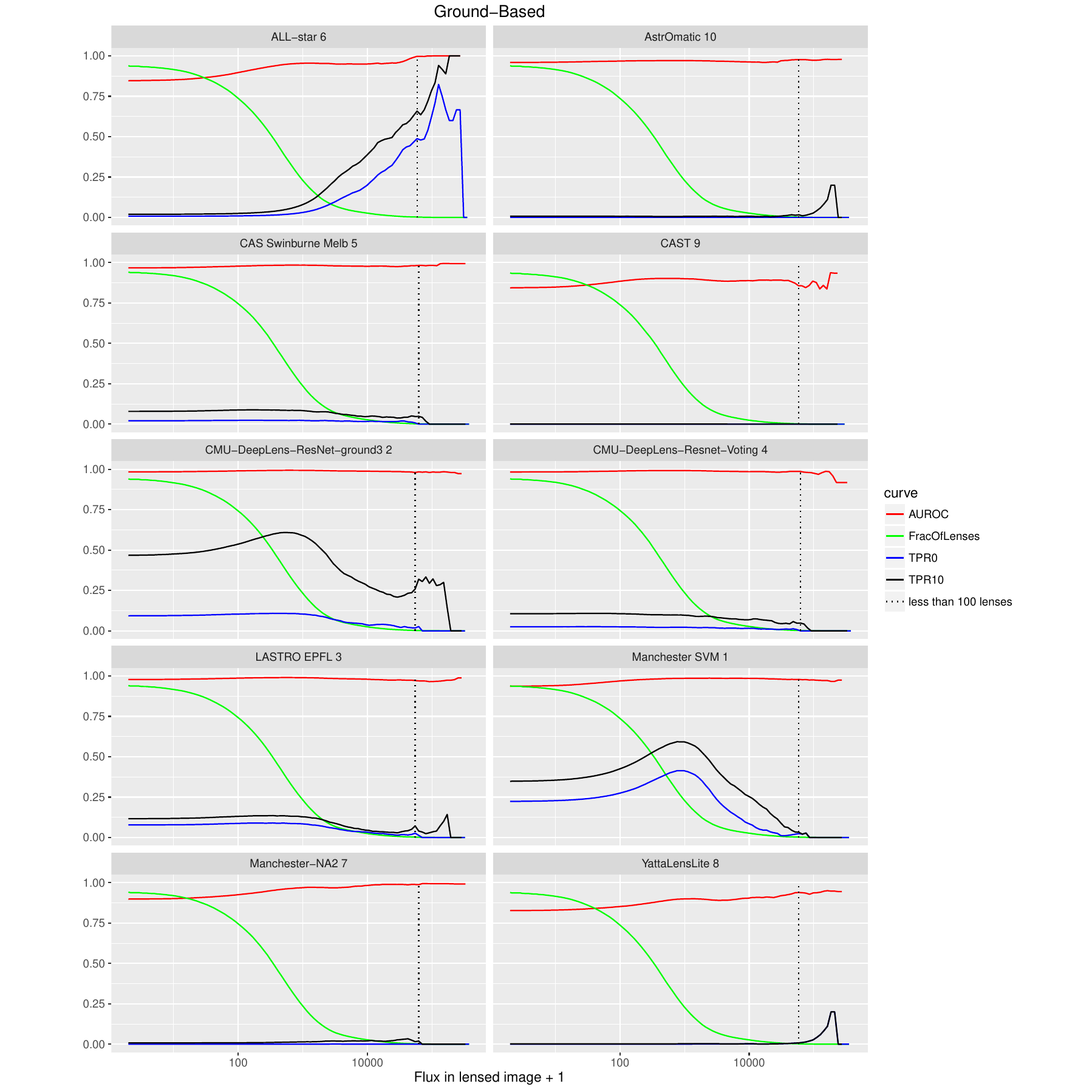}
 \caption{Same as figure~\ref{fig:flux_space}, but for ground-based entries.}
 \label{fig:flux_ground}
\end{figure*}

\begin{figure*}
\includegraphics[width=2\columnwidth]{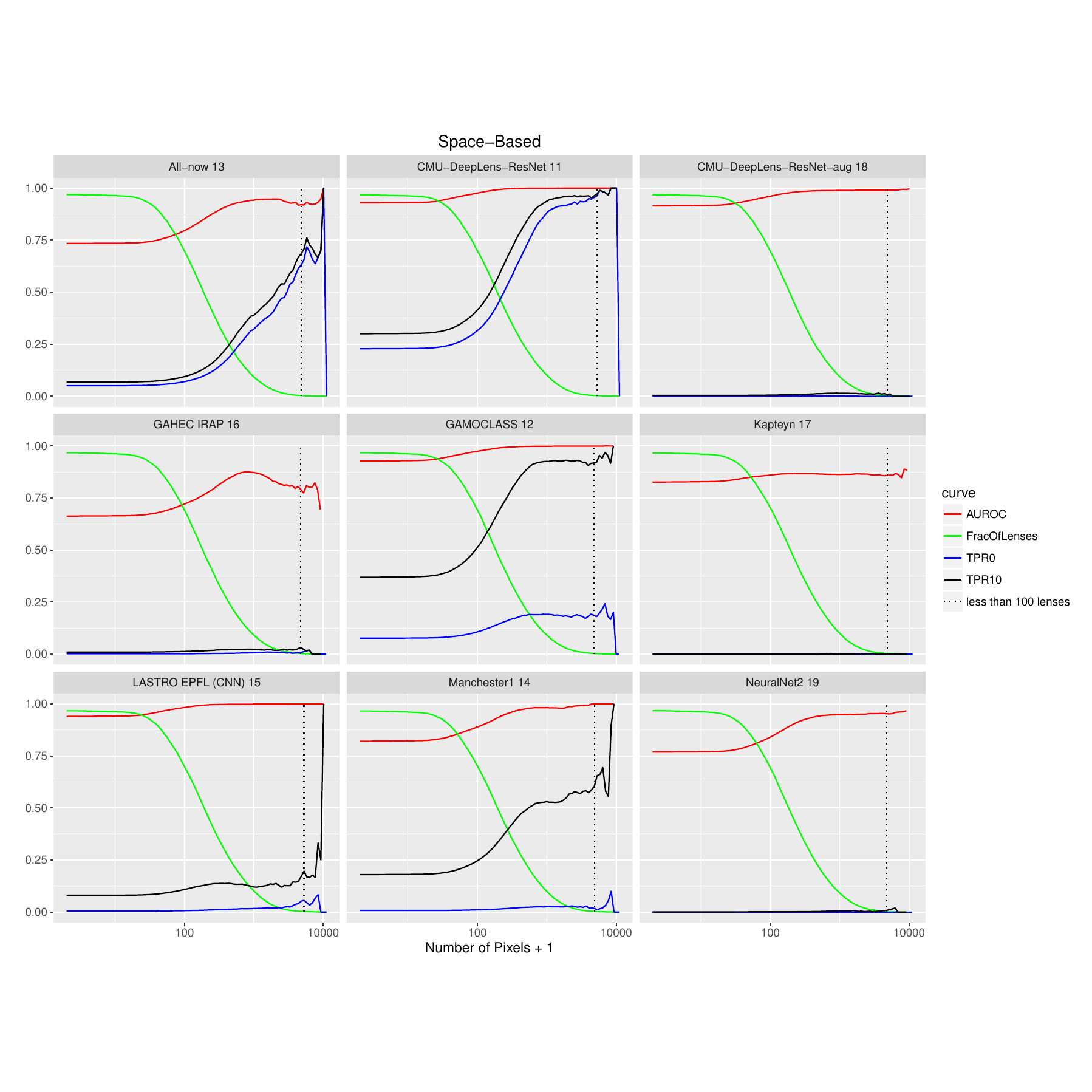}
 \caption{The AUROC, TPR$_0$, TPR$_{10}$ and the fraction of lenses as a function of the lens image size for the space-based test set.  The x-axis is the number of pixels that are above 1 $\sigma$ in the lensed source only image.  This is an indication of the lensed arcs' size.  The vertical dotted lines indicate where no more than 100 lenses in the test sample had larger Einstein radii. 
 }
 \label{fig:npixel_space}
\end{figure*}

\begin{figure*}
 \includegraphics[width=2\columnwidth]{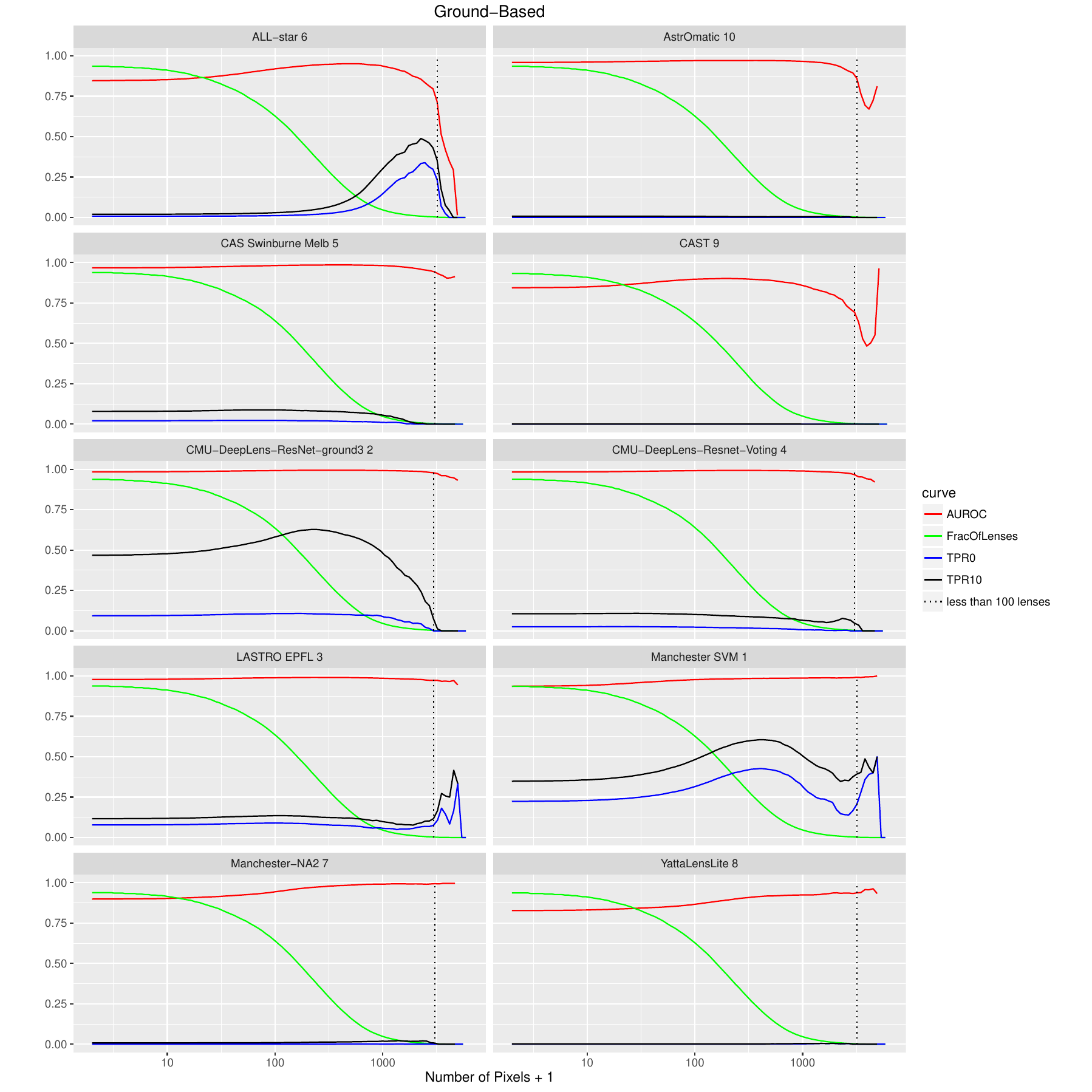}
 \caption{Same as figure~\ref{fig:npixel_space}, but for ground-based entries.}
 \label{fig:npixel_ground}
\end{figure*}

\begin{figure*}
 \includegraphics[width=2\columnwidth]{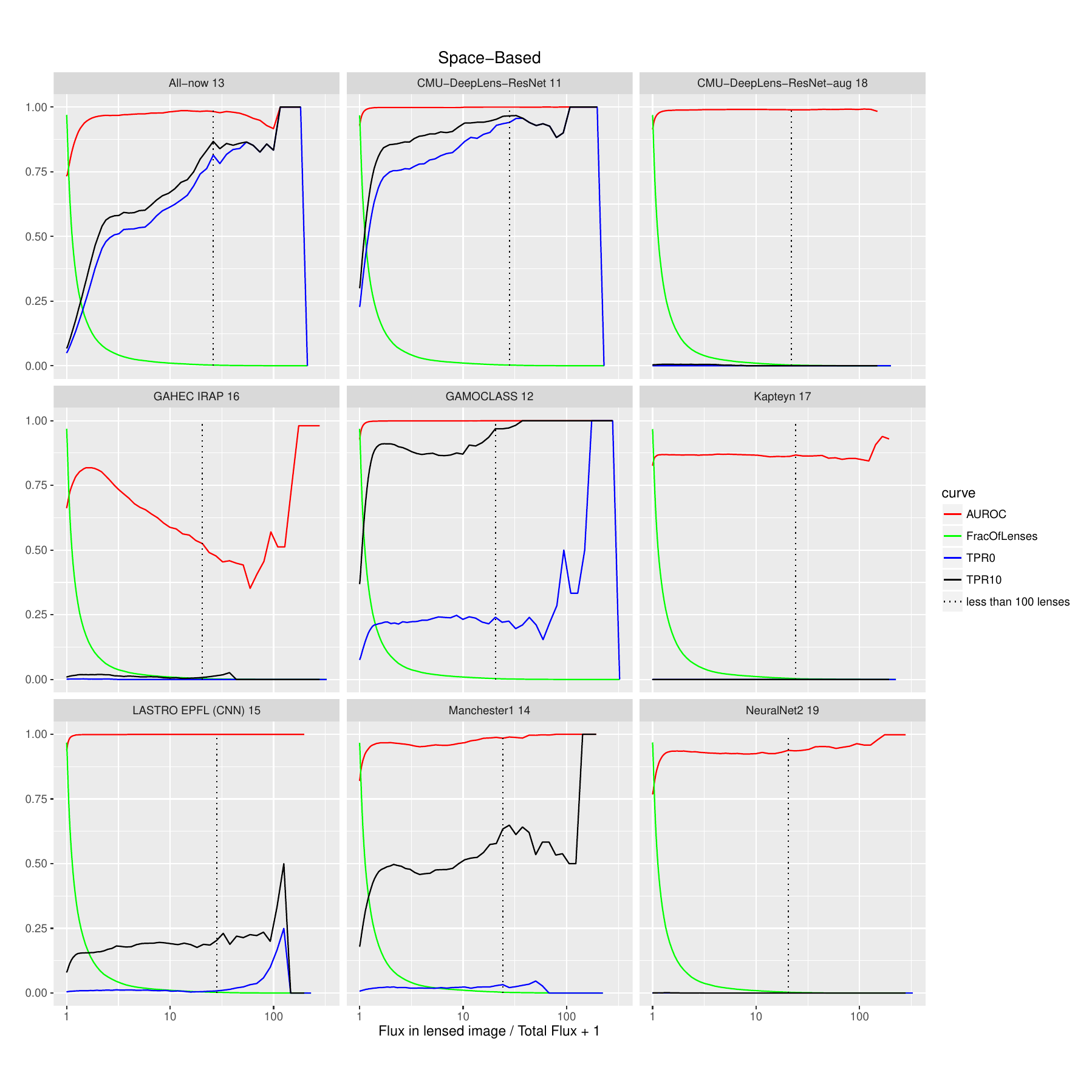}
  \caption{Same as figure~\ref{fig:flux_contrast_space}, but here the x-axis is the ratio of the flux coming from the lensed source to the total flux in the image in the index band.  This is for the space-based test set.}
 \label{fig:flux_contrast_space}
\end{figure*}

\begin{figure*}
 \includegraphics[width=2\columnwidth]{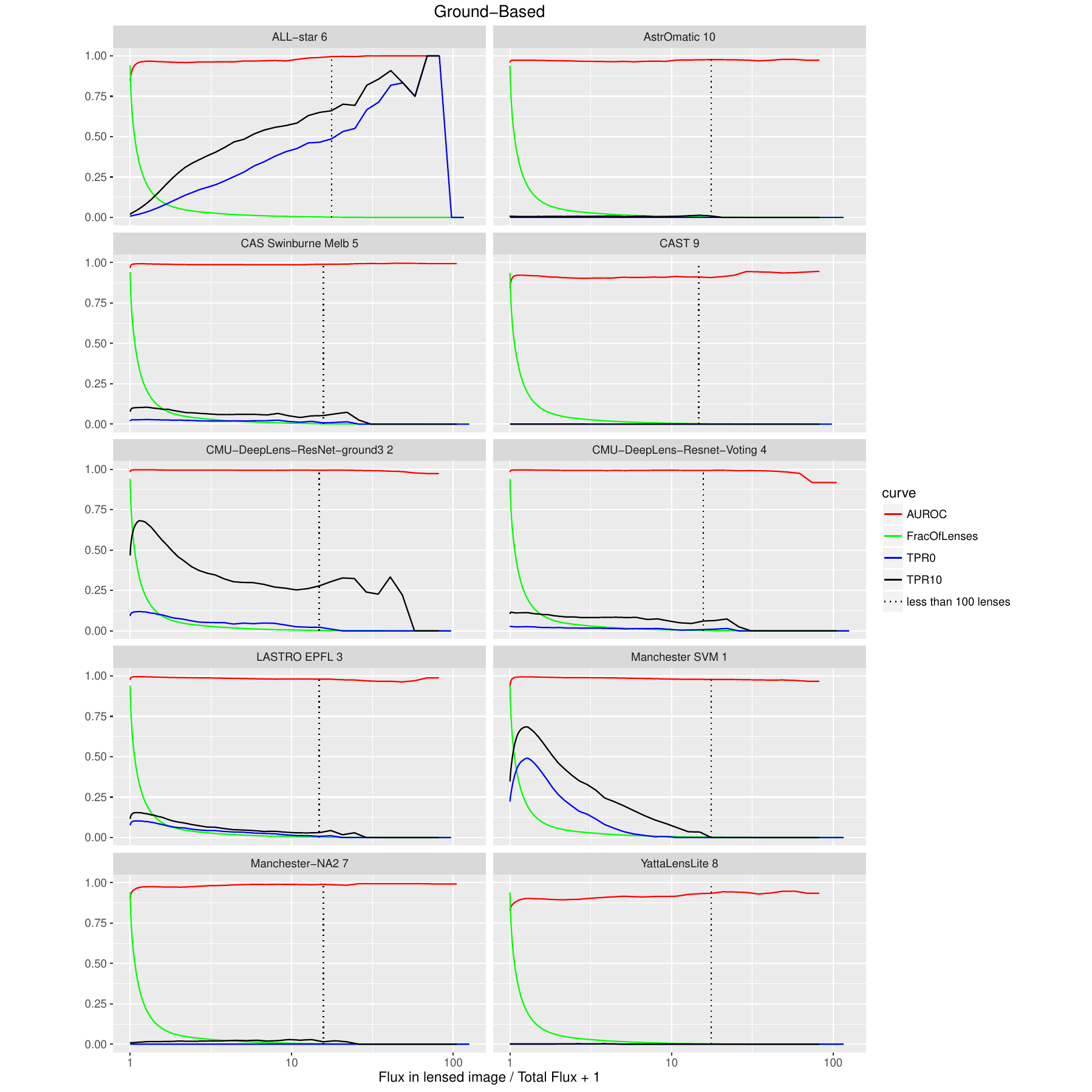}
 \caption{Same as figure~\ref{fig:flux_contrast_space}, but for ground-based entries.}
 \label{fig:flux_contrast_ground}
\end{figure*}

Figures~\ref{fig:einstein_space} and \ref{fig:einstein_ground} show the AUROC, TPR$_0$, TPR$_{10}$ and fraction of lenses as a function of a lower cutoff on the Einstein radius (area).  There is also a vertical dotted line that indicates 
where no more than 100 lenses in the test sample had larger Einstein radii.  Beyond this point one should be suspicious of small number statistics.  When deriving the distribution of Einstein radii from data these curves would need to be used to correct for detection bias.  It can be seen that CMU-DeepLens-Resnet, Manchester1, Manchester SVM and GAMOCLASS obtain significantly higher TPR$_0$ and  TPR$_{10}$ for larger Einstein radii.  Manchester1 is the human inspection method.  In some cases the TPR$_0$'s are above 50\% of the lenses that meet this criterion.  Remember that many of the so called lenses are very dim or there is no significant arc because the source position is well outside the caustic.  If an additional requirement was placed on the definition of a lens, such as the brightness of the arc being above a threshold, the TPRs would go up.

Figures~\ref{fig:flux_space} and \ref{fig:flux_ground} are the same except that the flux in the lensed images is used as the threshold.  We count only the flux in pixels with flux over one $\sigma$ of the background.  In some cases one can see an abrupt rise in the TPRs at some flux threshold.  CMU-DeepLens-Resnet in particular reaches a TPR$_0$ above 75\% for the brightest $\sim$ 10\% of the lenses. 

A lensed image can be bright without being visibly distorted as in the case of unresolved images.  Figures~\ref{fig:npixel_space} and \ref{fig:npixel_ground} use the number of pixels in the lensed image(s) that are over  one $\sigma$ of the background.  In this case also some classifiers show an abrupt improvement when the image is required to be larger than some threshold.  Interestingly in some cases the TPRs go down with lensed image size after reaching a peak.  This could be because they are not differentiating the arcs from companion galaxies as well in this regime.  There were also cases where the arc intersects with the borders of the image that might cause them to be missed.

Figures~\ref{fig:flux_contrast_space} and \ref{fig:flux_contrast_ground} investigate how the flux contrast between the foreground objects and the lensed source affects the classifiers.  Interestingly some methods' TPRs go up with this quantity and some go down.  We have not yet found any clear explanation for this variety of behaviours.  

The two human inspectors, NJ and AT, got significantly different scores on the ground based test set with individual AUROCs of 0.88 and 0.902 and TPR$_{10}$s of 0.01 and 0.06 respectively.  They did not inspect the same images however so differences cannot be considered conclusive, but it does suggest that different inspectors will have different detection efficiencies and biases.


\section{Conclusions and discussion}
\label{sec:conclusion}

A large variety of lens finding methods were tested on simulated images that were developed separately.    Based on figures~\ref{fig:einstein_space} and \ref{fig:einstein_ground}, we found that some methods could recover more than 50\% of the lenses above a lensed image brightness or size threshold without a single false positive out of 100,000 candidates.  If the data closely resembled the simulations we would already have reasonably good methods whose efficiency and biases can be quantitatively characterized.

We have done a fairly good job of determining that lenses can be identified in a population of fairly "normal" galaxies.  It is the rare "abnormal" objects that pose the greatest challenge.  When real KiDS data was used in the simulations the classifiers were all less accurate and it was only human inspection that found the one jackpot lens (a double Einstein ring with two background sources) in the data.  Things like ring galaxies, tidal tails in merging galaxies and irregular galaxies can be mistaken for lenses and were not well represented in the simulated data.  Accurately reproducing these objects will be an objective of future work.  This might be done by including more real images in the challenge or images based on real images with some random elements added.

It was surprising to some of the authors how well CNN and SVM methods did relative to human inspection.  These methods find differences in the classes of images that are not obvious to a human and can classify things as lenses with high confidence where a human would have doubt. 
This ability comes with some danger of over fitting to the training set however.  The distinguishing characteristics might only be a property of simulated data and not of real data.  In principle, SVM methods might potentially mitigate this somewhat because with them one can choose which features to use based on knowledge of the properties of irregular galaxies or ring galaxies for example.  This has yet to be shown however.  Methods based on fitting with a lens modelling code \citep{2009ApJ...694..924M,2017arXiv170401585S} might also help to mitigate this problem.
The confidence one will have in the machine learning methods is really limited by the confidence one has in the realism of the simulations.   It might be useful in the future to have a challenge without a training set.  This might more clearly reveal the presence of over-fitting.  
It would also be useful to include more real images or images more closely based on real images.

When initiating this project we had a concern that current methods would be too slow or require too much human intervention to handle large data sets.  Happily this seems not to be a problem with most of the automatic methods.  The CNN and SVM codes take some time to train, but once trained they are very fast in classifying objects.  Billions of objects can be easily handled.

Another lesson is that colour information is very important.  Even with lower noise levels, higher resolution, a simpler PSF and no masking, the lenses in the space-based set were harder to find than the lenses in the ground-based set (see figures~\ref{fig:roc_space} and \ref{fig:roc_ground}).  Having multiple bands clearly makes a significant difference.  Euclid will have several infrared bands with lower resolution than the visible images that were not included in the challenge.  Even rather low resolution information from another instrument or telescope when combined with higher resolution data in one band might significantly improve the detection rates.  Combining ground based data, such as LSST, with space based data, such as Euclid, would likely boost the detection rates by factors of several.

\section*{Acknowledgements}
AS was supported by World Premier International Research Center Initiative (WPI Initiative), MEXT, Japan.
RBM's research was partly part of project GLENCO, funded under the European Seventh Framework Programme, Ideas, Grant Agreement n. 259349. AT acknowledges receipt of an STFC postdoctoral research
assistantship.
We thank the International Space Science Institute (ISSI) for hosting and funding our workshop.\footnote{http://www.issibern.ch/}
JPK and CS acknowledge support from the ERC advanced grant LIDA and the ESA-NPI grant 491-2016.
CA acknowledges support from the Enrico Fermi
Institute at the University of Chicago, and the University of Chicago
Provost's Office. NL would like to thank the funding support from
NSFC, grant no.11503064, and Shanghai Natural Science Foundation,
grant no. 15ZR1446700. This work was also supported in part by the
Kavli Institute for Cosmological Physics at the University of Chicago
through grant NSF PHY-1125897 and an endowment from the Kavli
Foundation and its founder Fred Kavli. LVEK, CEP, CT and GV are supported through an NWO-VICI grant (project number 639.043.308). GVK acknowledges
financial support from the Netherlands Research School for
Astronomy (NOVA) and Target. Target is supported by
Samenwerkingsverband Noord Nederland, European fund for regional
development, Dutch Ministry of economic affairs, Pieken in de
Delta, Provinces of Groningen and Drenthe. 


\bibliographystyle{mnras}
\bibliography{references}


\appendix

\label{lastpage}
\end{document}